\documentclass[11pt,letter]{article}
\usepackage{cite}
\usepackage{amsmath,amssymb,amsfonts}
\usepackage{amsthm}
\usepackage{algorithm}
\usepackage{graphicx}
\usepackage{textcomp}
\def\BibTeX{{\rm B\kern-.05em{\sc i\kern-.025em b}\kern-.08em
    T\kern-.1667em\lower.7ex\hbox{E}\kern-.125emX}}
    
\usepackage[table]{xcolor}
\usepackage{caption}
\usepackage{braket}
\usepackage{bm}
\usepackage{booktabs}
\usepackage{pifont}

\usepackage{fullpage}

\usepackage{subfiles}

\usepackage{algpseudocode}
\usepackage{arydshln}
\newtheorem*{definition}{Definition}
\def\gs{g(\mathcal{S}, \alpha)}

\newtheorem{defn}{Definition}

\DeclareMathOperator*{\argmax}{arg\,max}

\title{Neural Architecture Search Algorithms for Quantum Autoencoders}

\usepackage{authblk}

\author[1,*]{Ankit Kulshrestha}
\author[1]{Xiaoyuan Liu}
\author[1]{Hayato Ushijima-Mwesigwa}
\author[2]{Ilya Safro}

\affil[1]{Fujitsu Research of America, Santa Clara, CA}
\affil[2]{Department of Computer and Information Sciences, University of Delaware, Newark, DE 19716, USA}
\affil[*]{Corresponding author: akulshrestha@fujitsu.com}

\begin{document}



\maketitle

\begin{abstract}
    The design of quantum circuits is currently driven by the specific objectives of the quantum algorithm in question. This approach thus relies on a significant manual effort by the quantum algorithm designer to design an appropriate circuit for the task. However this approach cannot scale to more complex quantum algorithms in the future without exponentially increasing the circuit design effort and introducing unwanted inductive biases. 
Motivated by this observation, we propose to automate the process of cicuit design by drawing inspiration from Neural Architecture Search (NAS). In this work, we propose two Quantum-NAS algorithms that aim to find efficient circuits given a particular quantum task. We choose quantum data compression as our driver quantum task and demonstrate the performance of our algorithms by finding efficient autoencoder designs that outperform baselines on three different tasks - quantum data denoising, classical data compression and pure quantum data compression. Our results indicate that quantum NAS algorithms can significantly alleviate the manual effort while delivering performant quantum circuits for any given task.







\end{abstract}




\section{Introduction}\label{sec:intro}


Quantum computation operates on a different method of processing and storing information compared to  classical computers. This difference stems from the fact that the information and algorithmic principles in quantum computation are based on those of quantum mechanics, rather than of classical mechanics. Ultimately, the objective is to use this approach to achieve  speed-ups in solving computationally challenging problems in such areas as finance \cite{herman2023quantum}, simulation \cite{childs2018toward}, and optimization \cite{ushijima2021multilevel}.
In the current regime of Noisy Intermediate Scale Quantum (NISQ)~\cite{preskill2018quantum}, quantum hardware has support for only a few hundred of qubits and is highly susceptible to a myriad of noise sources. A class of algorithms called Variational Quantum Algorithms (VQAs) aims to leverage classical computers to ``train" parameterized quantum circuits (e.g., using gradient descent-like process) for a given cost function. However, it has been noted in earlier studies~\cite{cerezo2019variational, cerezo2020cost, kulshrestha2022beinit} that parameterized quantum circuits themselves are susceptible to issues stemming from entanglement, barren plateaus and wire noise. Thus, in order for quantum computing to scale to large problem instances, several fundamental algorithms need to be developed that can overcome the present issues in quantum hardware and algorithms.

Machine learning is perfectly positioned to  facilitate the quantum computing by learning and generalizing from existing data. 
In recent times, several machine learning algorithms have been used to improve the performance of quantum algorithms~\cite{verdon2019learning, kulshrestha2023metaopt}. Moreover, quantum error correction (QEC) also has benefited from machine learning algorithms~\cite{convy2022machine, fu2023benchmarking}. The design of quantum circuits is a crucial step in quantum algorithm development. In most existing works, the design of quantum circuits is left to the algorithm designer and in turn induces several inductive biases in the overall pipeline. Some examples of inductive biases include the gate placement strategy, the number of layers, parameter initializing distribution (if dealing with parameterized circuits) and entanglement strategy. The presence of these inductive biases leads to ad-hoc designs that are often inefficient in terms of the number of actual resources required to solve the problem. We argue that automating the search process for a quantum circuit can overcome many of the inefficiencies induced by manual design and reduce the cost of deployment on costly quantum hardware. This motivates us to focus on developing Quantum Architecture Search algorithms in this paper.


In the current era of NISQ computers, depending on the architecture, the size of the problems are limited to a few tens or hundred qubits. Since a fault tolerance scheme that can effectively reduce the noise is some years away, our best hope of scaling quantum circuits to larger problems is by employing data compression. Reliable quantum data compression can help NISQ devices to explore problem sizes which are not efficiently classically simulable. Romero~\emph{et al.}~\cite{romero2017quantum} have introduced the idea of quantum autoencoder and have shown that it is possible to compress quantum data into fewer qubits. However, the ability to compress data is not enough. The encoder must be as efficient as possible since we want to induce very little overhead for quantum circuits solving the downstream task on a compressed input space. Motivated by this need, we focus our quantum architecture search algorithms on the problem of finding efficient encoding circuits for quantum data compression. We summarize our contributions as following:
\begin{itemize}
    \item We propose a novel design for modeling quantum circuits in quantum search and redefine the search problem from searching circuits to discovering repeatable \emph{patterns} that can be generalized across different tasks.

    \item We propose two novel neural architecture search algorithms that take advantage of our aforementioned formalism by combining evolutionary search and reinforcement learning techniques in an end-to-end fashion.

    \item We demonstrate the effectiveness of our algorithms in various combinations in hybrid quantum classical and pure quantum experiments. \emph{More specifically, we show that our algorithms yield parameter efficient quantum encoders that can compress information with high fidelity}. 
    
\end{itemize}

\section{Background and Related Work}\label{sec:bground}
Let us define the key terms and describe the quantum autoencoder (QAE) model in some detail.

\begin{defn}
    \textbf{Quantum Data Compression}: Let $\ket{\psi} \in \mathbb{C}^m$ be a quantum state in a given Hilbert space $\mathcal{H}$. The problem of quantum data compression is to find a state $\ket{\phi} \in \mathbb{C}^n$ where $n < m$ and $\ket{\phi} \in \mathcal{H}$ such that the fidelity $\mathcal{F}(\ket{\psi}, \hat{\ket{\psi}})$ is maximized. Here, $\hat{\ket{\psi}} = U(\bm{\theta})\ket{\phi}$ is the reconstructed state from $\ket{\phi}$.
\end{defn}

The problem of quantum data compression is similar to that in the classical domain. While it is possible to come up with the Huffman-like encoding scheme for quantum data, guaranteeing its reliability in capturing signal from noise is hard owing to the limitations of the NISQ computers. 
An alternative solution to this issue could involve formulating it as an optimization problem and training a model similar to an autoencoder to achieve a compressed representation of the input quantum state. In this paper, we build on the existing quantum autoencoder framework by Romero~\emph{et al}~\cite{romero2017quantum} (briefly discussed below) and focus on the problem of finding efficient encoder representation using neural architecture search techniques.


\begin{figure}[h]
    \centering
    \includegraphics[width=.5\columnwidth]{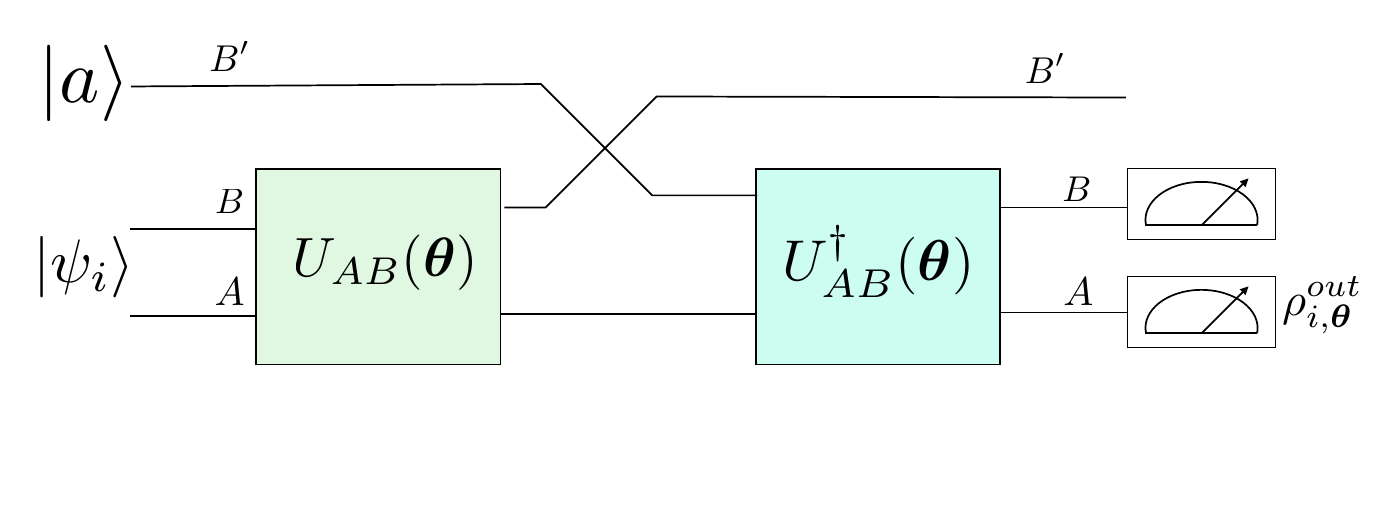}
    \caption{The QAE model diagram  presented by Romero~\emph{et al.}}
    \label{fig:qae_model}
\end{figure}

\textbf{Quantum Autoencoder: }  In the QAE model,  the input quantum state is interpreted to be a bi-partite state $\ket{\psi_{in}} \in \mathcal{H}^A \otimes \mathcal{H}^B$. The subsystem $A$ is called the latent system and the subsystem $B$ is called the \emph{trash} system. Essentially, we want to project the data contained in the original bi-partite state into a vector spanning only $\mathcal{H}^{A}$. The overall goal is to learn the parameters $\bm{\theta}$ of the quantum circuit in a way that minimizes the cost. 

A classical autoencoder is typically trained by minimizing the negative log likelihood between the decoded symbol and the input symbol. The analogous quantity in the quantum case for two mixed quantum states $\rho_{in}, \rho_{out}$ is the \emph{state-fidelity} $\mathcal{F}(\rho_{in}, \rho_{out})$ and is defined as: 

\begin{equation}\label{eq:state-fid}
    \mathcal{F}(\rho_{in}, \rho_{out}) = Tr[\sqrt{\sqrt{\rho_{in}}\rho_{out}\sqrt{\rho_{in}}}]^2
\end{equation}

A successful compression implies $\mathcal{F}(\rho_{in}, \rho_{out}) \approx 1$. In the QAE model, the decoder is the adjoint of the encoder unitary. Hence, learning the parameters of the encoder is sufficient to construct the entire autoencoder. A simple cost function for learning the parameters $\bm{\theta}$ for a given dataset of input states $\{p_i, \ket{\psi}_i \}_{i=1}^{K}$ can be:

\begin{equation}\label{eq:qae_vanilla}
    C_1(\bm{\theta}) = \sum_i p_i \mathcal{F}(\ket{\psi_i},\rho^i_{out}(\bm{\theta})),
\end{equation}
where $p_i$ is the associated probability of sampling the $i^{th}$ input state. We can write $\rho^i_{out}$ as:

\begin{equation}\label{eq:rho_pure_quantum}
\rho^i_{out} = U^\dagger_{AB}(\bm{\theta})Tr_B[\ket{\psi^i}\bra{\psi^i} U_{AB}(\bm{\theta})],
\end{equation}

\begin{figure*}[t]
    \centering
    \includegraphics[width=.75\columnwidth,height=3.2cm]{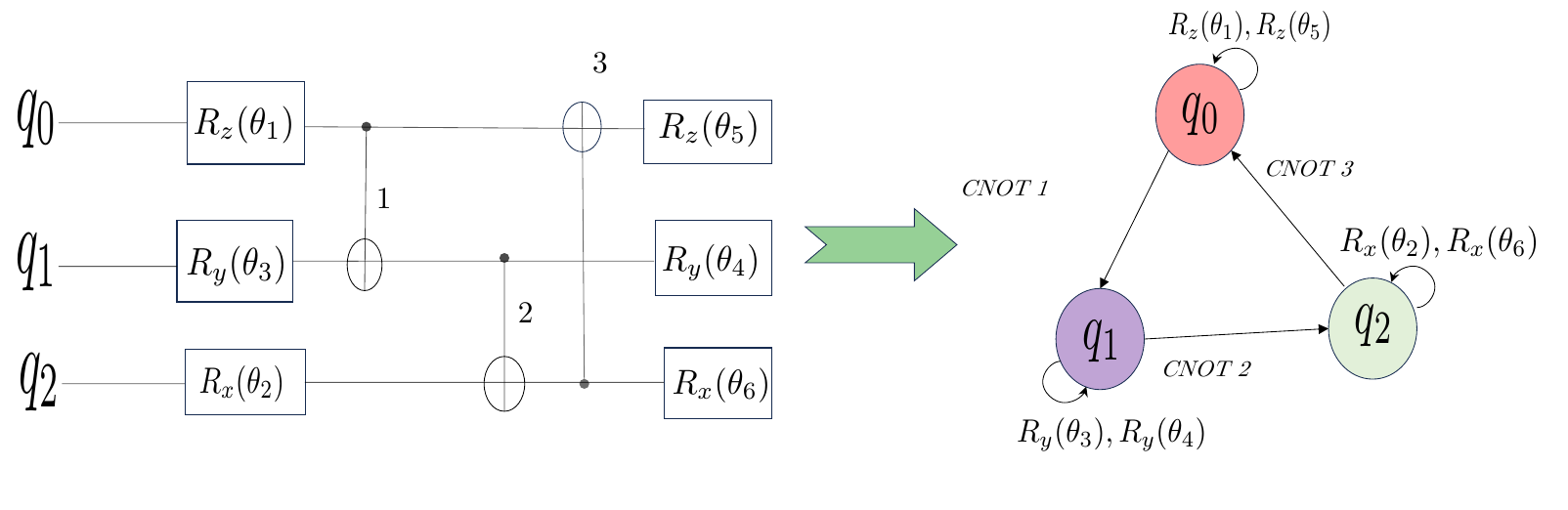}
    \caption{A compact representation of a logical quantum circuit. The self loops correspond to all single qubit operation on a given qubit node. A directed edge corresponds to all two qubit operations between two qubit nodes.
    }
    \label{fig:qckt_cell}
\end{figure*}

where $Tr_B[.]$ indicates a partial trace over the trash subsystem. Using this definition of $\rho_{out}$ we can easily compute $C_1(\bm{\theta})$. However, in practice the cost function proposed above requires preparing multiple \emph{identical} copies of the same input state to guarantee with high probability that the compressed state is indeed a true representation of the input quantum state. To overcome this problem, Romero~\emph{et al} propose a system demonstrated in Figure~\ref{fig:qae_model}. Here, a known reference subsystem $\ket{a}_{B'}$ is prepared and a swap test is performed between the trash and the reference subsystems. Additionally, instead of tracing out the trash subsystem, we trace out the latent subsystem and measure the fidelity after a swap test has been performed. This cost function is written as: 
\begin{equation}\label{eq:qae_task}
    C_2(\bm{\theta}) =  \sum_i p_i \mathcal{F}(Tr_A[U_{AB}(\bm{\theta})[\psi_{AB}]U^\dagger_{AB}(\bm{\theta})], \ket{a}_B'),
\end{equation}
where $[\psi_{AB}] = \ket{\psi}_{AB}\bra{\psi}_{AB}$. The ``traced-out" quantum state in the $B$ sub-system after applying the parameterized unitary $U_{AB}$ is $Tr_A[U_{AB}(\bm{\theta}) [\psi_{AB}]U^\dagger_{AB}]$. 
This cost function is not as expressive as the one in Equation~\ref{eq:qae_vanilla} but is immediately usable since one can easily prepare a known reference state. A variation of this cost function uses an average measurement across all trash qubits, since it has been shown that local measurements can help alleviate the notorious barren plateau problem in quantum circuits~\cite{cerezo2020cost} to some extent.

\section{Quantum Architecture Search for Quantum Autoencoders}\label{sec:qas_algs}



\begin{figure*}[htbp]
    \centering
    \includegraphics[width=.75\columnwidth]{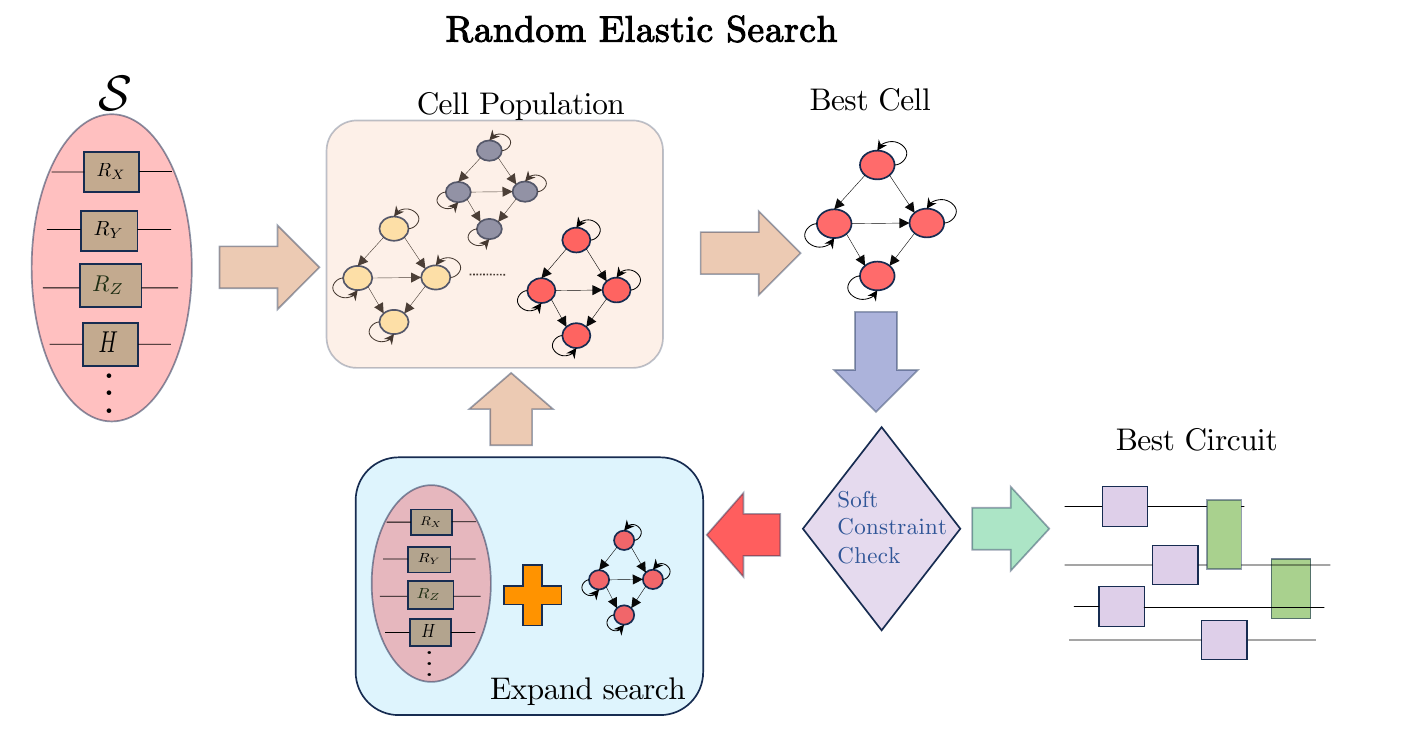}
    \caption{Random Elastic Search (RES) algorithm. A population of random cells is generated and then incrementally expanded with the best cell until a soft constraint is met.}
    \label{fig:res_flow}
\end{figure*}

\subsection{Quantum Architecture Search}\label{sec:qnas_description}

The field of neural architecture search (NAS) was born out of the necessity to design neural network architectures that could outperform hand-designed architectures on a given task. The goal was to reduce the development time and errors if a given task was well defined. Quantum Architecture Search (QAS) also shares the same motivations from its parent field and aims to find quantum circuits for various variational quantum algorithms that have been proposed in literature.

A QAS algorithm consists of three main components - a search space, search algorithm and a cost function. A search space $\mathcal{S}$ is a set of quantum gates over which the search algorithm performs its search. Some examples of search spaces are: (1) set of single qubit Clifford gates $\mathcal{S}_{Sc} = \{H, S, T, I\}$, (2) full Clifford group $\mathcal{S}_{C} = \{H, S, T, I, CNOT \}$ and (3) a generic set of quantum gates $\mathcal{S}_G = \{R_x, R_y, R_z, CNOT, CRY, CRZ, \dots \}$. The search space can also vary depending on the supported gates in a physical quantum computer.  We assume that we have a quantum circuit of the form:

\begin{equation*}\label{eq:pqc}
    U(\bm{\theta}) = \prod_{l=1}^L U_l(\bm{\theta}_l)
\end{equation*}

Where $U_l$ is a unitary operator representing a particular layer (i.e. a set of operations that can be performed in parallel in a single time step) and $\bm{\theta}_l$ represents the parameters of the various gates comprising that layer. We assume we have access to a generating function $f_{gen}(\mathcal{S})$ that allows us to generate a set of random unitaries $\mathcal{U} = \{ U_1, U_2, U_3, \dots U_M \}$ from the given $\mathcal{S}$. In this work, we reduce the task of finding the best performing circuit $U(\bm{\theta})$ to the task of finding the best layers $U_l$ for a given task:

\begin{equation}\label{eq:qnas_obj}
    U^b_l = \argmax_{\bm{\theta}^*_l}\{C(U_1), C(U_2), C(U_3), \dots C(U_M)\},
\end{equation}
where $C(U_i)$ corresponds to the cost function in Equation~\ref{eq:qae_task} and $U_i$ is shorthand for $U_i(\bm{\theta}_i)$. The best layer unitary is the one that minimizes the cost function the most with its optimized parameters. In works that consider QAS with this strategy~\cite{wu2023quantumdarts} (micro-search), it is typical to generate $U_b(\bm{\theta})$ as 

\begin{equation*}
    U^b(\bm{\theta}) = \prod_{l=1}^L U^b_l(\bm{\theta}^*_l), 
\end{equation*}
where $U^b_l$ is the best layer for a given task. and $\bm{\theta}^*_l$ correspond to the optimized parameters of the layer. Strategies of this type generally limit themselves to finding a single $U^b_l$ on a given cost function and then stacking them to an arbitrary depth to achieve competitive performance. 

On the other hand, we take a different approach with our proposed QAS algorithms. Instead of  performing a global search (i.e. directly search for $U^b(\bm{\theta})$) or a local search (i.e. search for one good layer and then repeat it to a given depth), we design an \emph{incremental} search procedure that interpolates between these two extremes. In order to achieve this, we introduce an ``expansion" phase in our search algorithm that builds a global circuit by incrementally searching for best layers and appending it to the previously best discovered ones. Let $\hat{U}^t_l$ denote the best layer at timestep $t$, then at timestep $t+1$, we find $\hat{U}^{(t+1)}_l$ as:

\begin{equation*}
    \hat{U}^{(t+1)}_l = \argmax_{\bm{\theta}^*_l}\{C(\Tilde{U}_1), C(\Tilde{U}_2), \dots C(\Tilde{U}_M) \},
\end{equation*}
where $\Tilde{U}_i = \hat{U}^t_l U_i$. Thus, at the $j^{th}$ iteration in our search procedure, we have a global circuit of the form 

\begin{equation*}
    U_b(\bm{\theta}) = \hat{U}_l^0\hat{U}_l^1\dots \hat{U}_l^j.
\end{equation*}

Performing such an incremental search can get computationally expensive if we do not control the number of iterations in the search algorithm. However, explicitly controlling the number of iterations can lead to suboptimal performance in a circuit. To overcome this paradox, we introduce a novel concept of ``soft constraints'' (Section \ref{sec:res}) that allow us to model secondary performance criteria in addition to the main cost function.

Incremental search can also be interpreted as an iterative improvement over some existing circuit to boost its performance. More precisely, we assume that we have access to some circuit $U_{gen}(\bm{\theta})$ that has been either sampled from $\mathcal{S}$ or generated from some prior search algorithm. The incremental search then uses some optimization method to improve the performance of this circuit on the given cost function by applying some in-place changes (e.g. sub-circuit substitution, gate fusion etc). In Section~\ref{sec:relm} we present an evolutionary search algorithm that performs the aforementioned optimization with a novel mutation operator. We model the mutation operator as a deep neural network that \emph{learns} the most optimal mutation policy $\pi_{mut}$ for mutating the circuits jointly with the overall evolutionary search algorithm. 

\subsection{Quantum Circuit Encoding}\label{sec:cell_repr}

In order to enable a search algorithm to perform search over multiple possible architectures, an encoding is needed to represent them within the search space. Many different encodings like list based~\cite{altares2021automatic, zhang2021neural}, DAG based~\cite{duong2022quantum, nguyen2022quantum} and tensor based~\cite{patel2024curriculum} have been proposed in literature. Readers are encouraged to follow the survey by Martyniuk~\emph{et al.}~\cite{martyniuk2024quantum} and the references contained therein for a more comprehensive review of the encoding strategies used in other QAS algorithms.

Given that many different encoding strategies have been tried in recent literature, a natural question to ask is - what is the need for a newer type of encoding? There are two answers to this question. First, existing QAS methods map the problem of finding optimal circuits to the proximal problem of finding the best gate order that leads to a good performance on a given task. On the other hand, we propose to tackle a harder instance of the problem - we jointly aim to find the best connectivity as well as the best sequence of operations that not only lead to a good performance on the given task but also satisfy additional constraints placed on the search. Second, our proposed algorithms require an encoding that allows them to perform the expansion and selection phases efficiently. Our analysis of the current set of encoding methods led us to conclude that we needed a different type of encoding to power our search algorithms.


Figure~\ref{fig:qckt_cell} shows our proposed encoding. In our method, we represent a circuit as a directed graph with each node representing a single qubit in the circuit. This representation is not a strict one to one mapping, but rather an abstraction designed to tackle the aforementioned problem. A self-loop indicates single qubit operation on that particular qubit. The ``weight" on the self-loop edge is a loosely ordered sequence of all single qubit operations that may exist on that particular qubit. We ignore the strict ordering on qubits because we are operating in the space of logical quantum circuits where gate scheduling is irrelevant. An edge between two qubit nodes indicates a multi-qubit operation between the two nodes. The edge also admits vector valued weights where each component indicates a particular kind of multi-qubit operation. For example, in Figure~\ref{fig:qckt_cell}  if there were two CNOT operations between qubit 0 and qubit 1, they can be easily represented as a two dimensional weight vector over the edge connecting the qubits. When circuits are represented in this way, then it becomes possible to find structure since a new structure can be formed by simply adding or removing an edge between two nodes, while the sequence of operations can be obtained by simply changing the weight vectors. 

\subsection{RES Algorithm: Random Elastic Search}\label{sec:res}

We now introduce our first algorithm called Random Elastic Search (RES). It's overall algorithmic flow is shown in Figure~\ref{fig:res_flow}.

 
The RES algorithm is divided into two main phases - search and expansion. In the search phase, we sample $P$ randomly generated cells from $\mathcal{S}$ with the restriction that only a single two-qubit operation can be chosen per non self-loop edge (Here, $P$ is a hyperparameter indicating the number of cells to sample). Each of these cells correspond to a potential quantum circuit. We evaluate all cells in the population and select the cell that yields the best performance on the desired metric. For the efficiency reasons, the cells are lazily evaluated, i.e., the cells are not converted to full circuits until they are being trained.

 Once the best cell from the generated population is selected, the expansion phase begins. In this phase, we accept a user defined soft constraint $\mathcal{C}$. This soft constraint is a secondary constraint that we wish to enforce in the search process. For instance, if a user desires to limit the total number of parameters in the circuit to below a certain number, they can add it as a soft constraint. The soft constraint is checked at the beginning of the expansion phase. If the soft constraint is not met, we use this best cell as the \emph{seed} architecture for generating a new population from the search space. In other words, a cell in the new population consists of all operations in the seed cell and has randomly sampled additional operations from the search space. 

 These two phases alternate until the soft constraint is satisfied. When that happens, the best cell from the current population is converted to a circuit and returned to the user. A key advantage of this method is that it combines the simplicity of random search with a novel incremental search procedure that can satisfy additional constraints. Moreover, this algorithm can be used as a population initialization step in other quantum architecture search algorithms.

 Different from other micro-search strategies in literature~\cite{wu2023quantumdarts} where the ansatz is built by stacking an optimal ``layer" up-to a certain depth, RES finds optimal architectures by performing an expanding micro-search that incrementally builds the ansatz by conditioning the new best layer on previously discovered ones. This type of search has the advantage of finding non-intuitive layers that are efficient in terms of the given soft constraint.

 \subsubsection{Modeling Soft Constraints in RES}
 One of the key aspects of the RES algorithm is the ability to model ``soft constraints" in search. We have touched upon this topic above to help explain the overall algorithm. We now make this notion more precise: 

\begin{definition}
\textbf{Soft Constraints in NAS}: Given a search space $\mathcal{S}$, some discrete or continuous set of actions $\alpha$, a soft constraint is a function $\gs$ that penalizes the search algorithm based on given (upper or lower) bound on a quantity that is not directly optimized during search. 
\end{definition}

In many QAS algorithms, these soft constraints provide extra ability to the search algorithm to generate architectures that satisfy some desired properties. For example, in reinforcement learning (RL) based approaches for QNAS, the illegal action (IA) penalty has been proposed to weed out actions that lead to suboptimal circuits~\cite{patel2024curriculum}. In this case we can think of $\gs$  as $R(\tau) + \beta \gs$ where $\beta$ is some balancing coefficient. Similarly, in Genetic Algorithm (GA) based approaches there are often multiple objectives~\cite{chivilikhin2020mog} to model competing objectives of obtaining the best circuit with less number of two qubit gates. Here $\gs$ be expressed as $F(\mathcal{S}, \alpha, \mathcal{P}) + \gs$ where $F$ is the fitness function over a given population $\mathcal{P}$. 

One of the main drawbacks of this approach is that training an objective function with a constraint is quite computationally demanding. The problem is exacerbated by the fact that in noisy environments the only way to establish the performance of a quantum circuit is to run it on a simulator with an appropriate noise model. Our key contribution is to realize a simple function of the form $\gs < \tau$ where $\tau$ is the upper bound on the quantity we wish to optimize concurrent with the main objective. These ``soft-quantities" can include the total number of layers, total number of parameters, number of two qubit gates etc. This function type may appear simple, but it has the advantage of rendering our proposed search algorithm highly parallelizable. The function evaluations of $\gs$ are simply performed after all parallel evaluations in a particular iteration are finished. We illustrate soft constraints with an example: 

\textbf{Example} Assume a task of unitary compilation where we need to approximate an unknown target unitary $U_{target}$ within some precision distance $\epsilon$. One of the ways to do this would be to measure the fidelity between the evolved state $\mathcal{F}(\ket{\psi}_{target}, \ket{\psi(\theta)})$ where $\ket{\psi(\theta)} = U(\theta)|0\rangle$ and $U(\theta)$ is a generated unitary from our algorithm. If we wish to constrain the number of parameters to a certain limit $P$, we can model $\gs \simeq |\bm{\theta}| \leq P$. Our search procedure will now return circuits that have high fidelity \emph{and} and within the parameter budget.

        

\begin{figure*}
    \centering
    \includegraphics[width=.75\columnwidth]{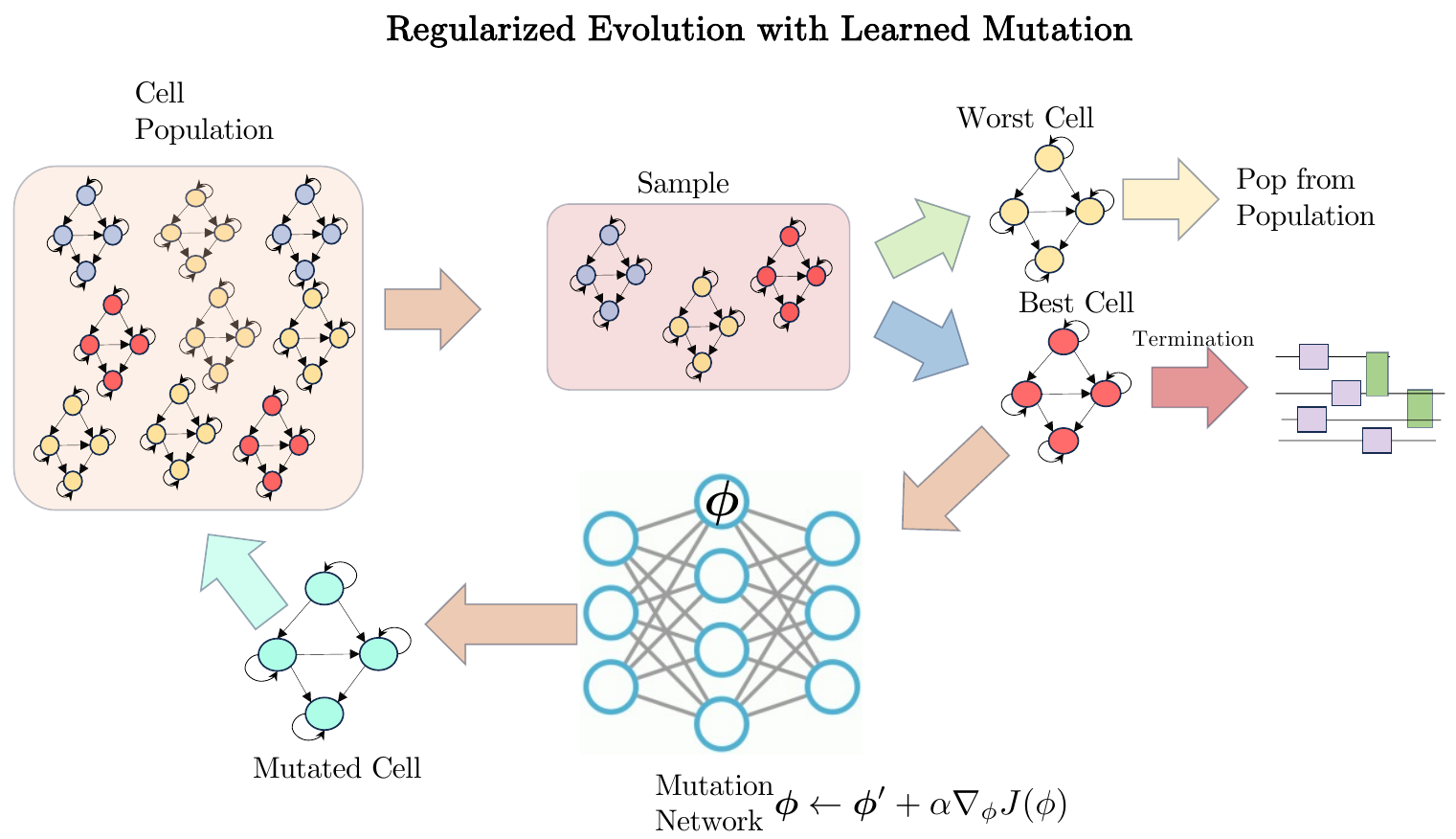}
    \caption{Regularized Evolution with Learned Mutation(RELM) algorithm. The algorithm combines the ideas of regularized evolution and learns a \emph{mutation} policy for mutating the best performing cells to search for best performing patterns.}
    \label{fig:relm_flow}
\end{figure*}

\subsection{RELM Algorithm: Regularized Evolution with Learned Mutation}\label{sec:relm}

One of the most common methods of architecture search is to learn a parameterized policy $\pi_\phi$ that can generate good candidate architectures. This policy is typically learned by reinforcement learning methods~\cite{sutton2018reinforcement}. However, a drawback of using reinforcement learning in architecture search is that the search process even for the most simple quantum circuits is extremely slow to start yielding profitable architectures~\cite{chen2019renas}. Evolutionary search techniques~\cite{real2017large, real2019regularized, goldberg1991comparative} on the other hand have been used before reinforcement learning and are able to find global solutions to search problems. 


In our proposed algorithm, we aim to combine the best characteristics of reinforcement learning and evolutionary search in a single end-to-end procedure. We build on the  ideas from Regularized Evolution~\cite{real2019regularized} and learned mutation in NAS~\cite{chen2019renas} applying them in the quantum domain. To the best of our knowledge, our work is the first one to consider a joint approach for quantum architecture search. 

Our novel algorithmic scheme is shown in Figure~\ref{fig:relm_flow}. There are three distinct phases of the algorithm - initialization, evolution and mutation. The initialization phase is concerned with coming up with a candidate initial population for evolution. In the typical evolutionary search algorithms including regularized evolution, this phase is a simple random search over possible architectures. While we allow for a similar initialization, our experiments show that we achieve better results when we use the preceding RES algorithm as the population initializer. A second advantage of using RES as a population initializer is that the secondary constraint can be followed with evolution and mutation phases.

During the evolution phase, we sample $S$ cells from the (initial) population. These sampled cells are then run on the target application and the best and worst performing cells are selected from amongst them. The worst performing cell is removed from the population while the best cell is selected for the mutation phase. We denote $E$ as the number of iterations for the mutation phase. Before proceeding to the mutation phase, we check if the number of iterations are complete and if so, we return the convert the current best cell to a circuit and return it to the user.


\begin{algorithm}[tb]
\caption{Mutation Pipeline in RELM}
\label{alg:mut_pipeline}
\begin{algorithmic}
    \Require Best cell: $c$, Mutation controller parameters: $\bm{\phi}$

    \State $\bm{S} \gets $ \Call{SINGLE\_QUBIT\_OPERATION}{$c$}
    \State $\bm{M} \gets $ \Call{MULTI\_QUBIT\_OPERATION}{$c$}
    \State $\mathcal{S}_{emb} \gets$ \Call{SINGLE\_OP\_EMBED}{$\bm{S}$}
    \State $\mathcal{M}_{emb} \gets$ \Call{MULTI\_OP\_EMBED}{$\bm{M}$}
    \State $J_{emb} \gets$ \Call{CONCATENATE}{$\mathcal{S}_{emb}, \mathcal{M}_{emb}$} \Comment{Input Embeddings}

    \State $\mathcal{A} \gets \mathcal{E}^{N}(J_{emb})$ \Comment{Transformer Encoders repreated $N$ times.}

    \State $\mathcal{O} \gets$ OUT\_PROJECTION($\mathcal{A}$)
    \State $\mathcal{S}_{out}, \mathcal{M}_{out} \gets$ SPLIT($\mathcal{O}$) \Comment{Output Projection}

    \State \Return $\mathcal{S}_{out}, \mathcal{M}_{out}$
   
\end{algorithmic}
\end{algorithm}

If the previous step does not terminate, we begin the mutation phase. The goal of this phase is to learn a mutation policy $\pi_\phi$ such that the child cell performs better than the parent cell. However, the mapping from a cell representing a quantum circuit to an input that can be used by a policy network is not straightforward. Hence, we engineer a novel mutation pipeline shown in Algorithm~\ref{alg:mut_pipeline}. The mutation controller consists of three sub-components: input embedding, attention module and the output embedding. We extract the single and multi-qubit operations from the given input cell and instantiate embedding matrices $W_{s} \in \mathbb{R}^{|S| \times e}$ and $W_{m} \in \mathbb{R}^{|M| \times e}$, where $|S|$ is the number of the single qubit operations, $|M|$ is the number of multi-qubit operations and $e$ is the embedding dimension for a single operation. The embeddings are then concatenated into a joint embedding and are input to the transformer encoder~\cite{vaswani2017attention}. In our experiments, we set the number of encoders $N=2$.  The output from the encoder is then fed to output projection $W_{out} \in \mathbb{R}^{e \times o}$, where $o$ is the output embedding dimension. For ease of training, we set $o = 2e$. This output embedding is then split into its respective rotation and entanglement operation embeddings for selecting the respective mutated operations. The parameters of the mutation controller are trained using the policy gradient loss~\cite{sutton1999policy}:
\begin{align}\label{eq:vpg_loss}
    \phi_{t+1} &= \phi_t + \alpha \nabla_\phi J(\bm{\phi})\nonumber\\
    \nabla_\phi J(\bm{\phi}) &= E_{\tau \sim \pi_\phi}[\sum_{t=0}^T \nabla_\phi log \pi_{\phi}(a_t | s_t) R(\tau)]
\end{align}
To encourage the policy network to generate architectures that perform better than the parent architectures we use the following reward function:
\begin{equation}\label{eq:relm_reward}
R(\tau) = 
\begin{cases}
    \mathcal{F}_{parent} - \mathcal{F}_{child} & \text{if}\ \mathcal{F}_{parent} > \mathcal{F}_{child}\\
    tan(\mathcal{F}_{child}\frac{\pi}{2}) & \text{otherwise}
\end{cases}
\end{equation}
Here, the first case corresponds to a \emph{negative} reward if the mutated child has a lower desired metric $\mathcal{F}$ than the parent. Otherwise, we use the reward function used by Chen~\emph{et al}~\cite{chen2019renas} since it shows a good performance in our experiments. We leave the design of a better reward function as a future work.

\section{Experiments}\label{sec:expts}

In this section, we examine the performance of our proposed search algorithms by conducting several experiments in different settings. For the RELM algorithm, we use the embedding dimension $e=32$ and the hidden dimension of the transformer encoder $h=64$. The evolution and mutation phase are run for $30$ epochs. For both algorithms, we set the initial population size $P = 30$. For each experiment, we evaluate the performance of our algorithms in three ways - standalone RES, RELM with the initial population initialized by a random search (RS) procedure and RELM with the initial population initialized by RES procedure.



\begin{figure*}
    \centering
    \includegraphics[width=\columnwidth]{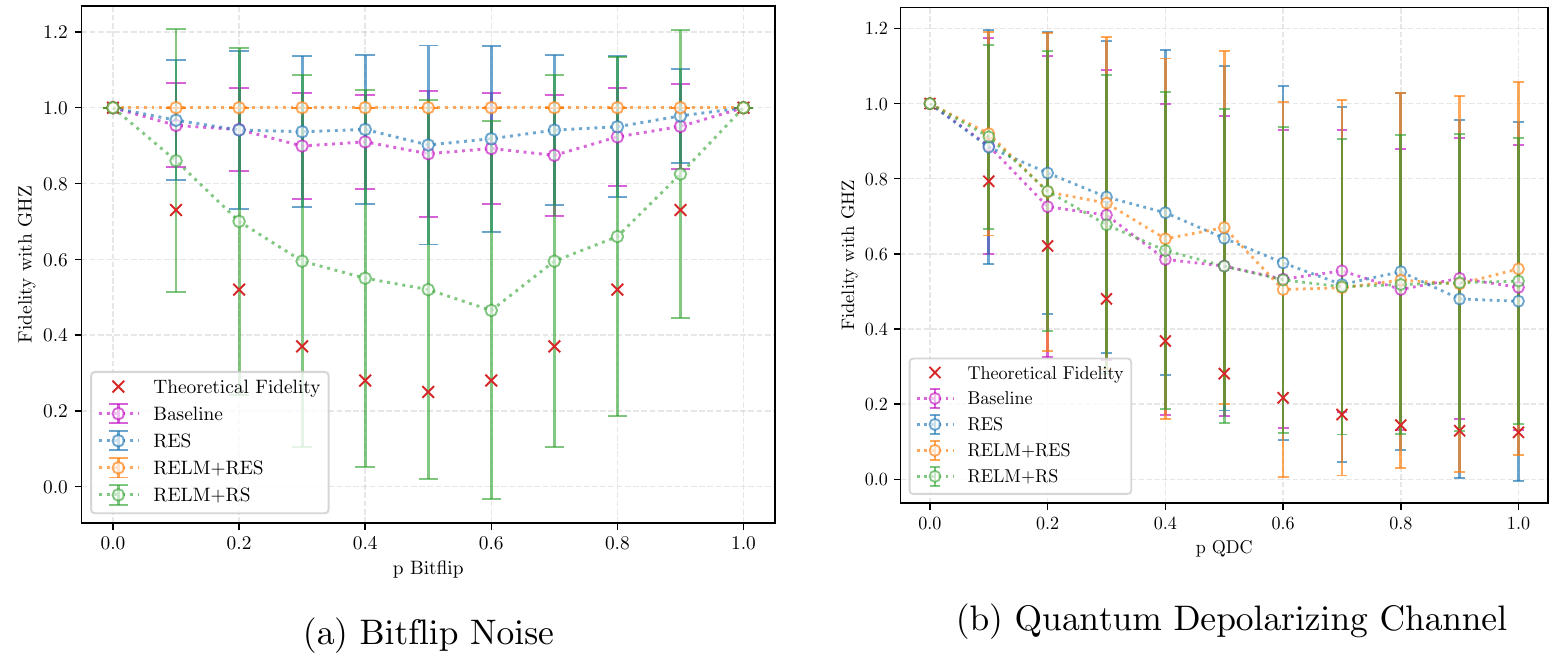}
    \caption{Performance of discovered circuits vs baseline on the test set with different noise types for a 3-1-3 QAE. Each point for every value of $p$ is an average over the 200 test samples and the error bars indicate the standard deviation.}
    \label{fig:denoising_results}
\end{figure*}
\subsection{Quantum Data Denoising}

Quantum autoencoders, like their classical counterparts have multiple applications. For instance, they have been successfully applied to denoise quantum data~\cite{bondarenko2020quantum}. In our first set of experiments we examine denoising performance of a quantum autoencoder with a twofold objective. First, we want to find autoencoder circuits that are efficient in depth or number of parameters. Second, we want their performance to either match or outperform a finely tuned baseline implementation. 

In our experiments we focus on denoising a 3-qubit GHZ state $|GHZ\rangle^{3} = \frac{\ket{0}^{\otimes 3}  + \ket{1}^{\otimes 3}}{\sqrt{2}}$ by finding an efficient encoding circuit for a [3-1-3] QAE. For constructing a noisy version of the GHZ state, we consider the same type of noise channels as proposed in a related work~\cite{achache2020denoising}: a bitflip channel and a Quantum Depolarizing Channel(QDC). In the former, a bit in the input quantum state is flipped with a finite $p_{bitflip}$. In the latter, a noisy state is obtained with a finite $p_{dep}$ as:

\begin{equation}\label{eq:qdc}
    \rho \rightarrow (1 - p_{dep})\rho + p_{dep}(\frac{1}{n}.\bm{1})
\end{equation}

Where $\bm{1}$ is the maximally mixed density matrix. It has been shown in~\cite{achache2020denoising} that this channel is equivalent to applying a Pauli matrix with probability $p_{dep} \in \{1-\frac{3p_{dep}}{4}, \frac{p_{dep}}{4}, \frac{p_{dep}}{4}, \frac{p_{dep}}{4}\}$ to each qubit. For each type of noise channel we create 100 training and validation circuits. Each noisy circuit is then applied to a 3-qubit $\ket{0}$ state and the resulting noisy state is input to the encoder circuit. All circuits are trained with $p_{noise}=0.2$ using the SWAP test and the cost function in Equation~\ref{eq:qae_task}. The training is done with Qiskit~\cite{cross2018ibm} with the COBYLA optimizer. The validation set is used to evaluate the performance of a proposed autoencoder circuit with tuned parameters. This evaluation ultimately guides the search algorithm in proposing better circuits. Once the best circuit is found we evaluate it on a test set comprising of test circuits constructed with $p_{noise} \in [0, 1]$. For each value of $p_{noise}$ we create 200 circuits and evaluate the fidelity with a clean state.

\begin{table}[h]
    \centering
    \begin{tabular}{c c c}
    \toprule
       Algorithm  & Bitflip & QDC \\
       \midrule
        RES & 21 & 20 \\
        RELM w/ RES & 16 & 7\\
        RELM w/ RS & 7 & 7 \\ 
        \hdashline
        Baseline & 48 & 48\\
        \bottomrule
    \end{tabular}
    \caption{Number of parameters in the circuits discovered by the search algorithms as compared against the best performing baseline.}
    \label{tab:denoising_comparison}
\end{table}

The baseline autoencoder circuit is carefully constructed from the blueprint provided by Romero~\emph{et al}~\cite{romero2017quantum} in their Circuit B. Through a grid search over various single and multi-qubit values we found a circuit with RZ single qubit rotation and controlled-RX entanglement to work best. We then sampled the parameters with different random seeds and chose the best performing set from amongst them as our baseline. In the case of RES algorithm, we restricted the search to finding circuits with less than 3 layers. This constraint was made stricter (i.e. maximum of 2 layers) when we used the RES algorithm as an initializer for the RELM procedure. The denoising results are reported in Figure~\ref{fig:denoising_results}. Note that $\Bar{\mathcal{F}} + \Delta \mathcal{F}$ can be greater than one. The number of parameters in the best circuits discovered by each algorithm for different types of noise are shown in Table~\ref{tab:denoising_comparison}.

From Figure~\ref{fig:denoising_results}a we can see that the RES algorithm finds a circuit that marginally outperforms the baseline. However, the RELM algorithm with RES initialization significantly outperforms the baseline. Indeed, for a noisy 3-qubit GHZ, the QAE discovered by RELM is able to \emph{perfectly denoise} the state irrespective of the amount of bitflip noise. With the random search (RS) initialization however, a sub-optimal circuit is discovered that performs worse than the baseline. We attribute this to the high variance in the random search initialization. The quantum depolarizing channel noise is a more stronger source of noise and thus is much harder to denoise for a QAE. Even so, from Figure~\ref{fig:denoising_results}b it is evident that the algorithms are able to denoise the noisy quantum state to a certain extent. Moreover, in the low noise threshold $p \in [0.0 , 0.2]$, we can see that the circuits returned by our search algorithms have a higher degree of denoising with lesser variance as compared to the baseline circuit. We show the aggregated performance over all values of $P$ for the algorithms in the Supplementary material.

\subsection{Downstream Quantum Compression}

To demonstrate the effectiveness of our algorithms in this setting, we build on the empirical setup of Bilkis~\emph{et al}~\cite{bilkis2021semi} and generate a dataset of 16 states corresponding to the ground state of a VQE solver for a $H_2$ molecule with different bond lengths. Similar to the previous setup, we train on six states and run validation on remaining ten states. Our goal is to compress the resulting 4-qubit state into a 2-qubit state with high fidelity. If we can successfully compress information in a quantum state into a lower dimensional quantum state with high fidelity, then we can scale the size of problems that can be handled in the aforementioned applications.Figure~\ref{fig:vqe_setup} shows the overall empirical setup for our experiments. For the input and output states we measure the log fidelity  ($-log_{10}(1-\mathcal{F})$) where $\mathcal{F}$ is the fidelity in Equation~\ref{eq:state-fid}  averaged over 10 independent runs over the training and test sets. For comparison, we report the results by Bilkis~\emph{et al.} on the same training and test set in terms of the logfidelity.

\begin{table}[h]
    \centering
    \begin{small}
    \begin{tabular}{p{5.25em}p{3.5em}cc}
    \toprule
      Algorithm  & Parameters & Train Fidelity & Test Fidelity\\
      \midrule
       RES  & 21 &  $5.62 \pm 1.09$ & $6.03 \pm 0.177$ \\ 
       RELM+RS & 10 & $6.01 \pm 3.04$& $4.79 \pm 3.42$\\ 
       RELM+RES & 12 & $6.76 \pm 0.45$ & $\bm{6.52}\pm 0.76$ \\
       \hdashline
       VAns~\cite{bilkis2021semi} & 45 & 6.49 &  6.17\\
       \bottomrule
    \end{tabular}
    \end{small}

    \caption{Performance of proposed algorithms on the training and test splits of the VQE dataset. We report the logfidelities as an average of 10 indepdendent evaluations of the trained autoencoder on the test set.}
    \label{tab:vqe_experiment_results}
\end{table}

Table~\ref{tab:vqe_experiment_results} shows the results of our proposed algorithm when benchmarked against the algorithm proposed by Bilkis~\emph{et al}. We can see that all our proposed algorithms get comparable or better results on the training and test sets with significantly fewer number of parameters than the benchmark unitary. More specifically, we find that the RELM algorithm with RES initialization yields the most parameter efficient and highest performing circuit that compresses the input quantum state with the highest fidelity (difference measured in logfidelity).

\begin{figure*}[t]
    \centering
    \includegraphics[width=\columnwidth]{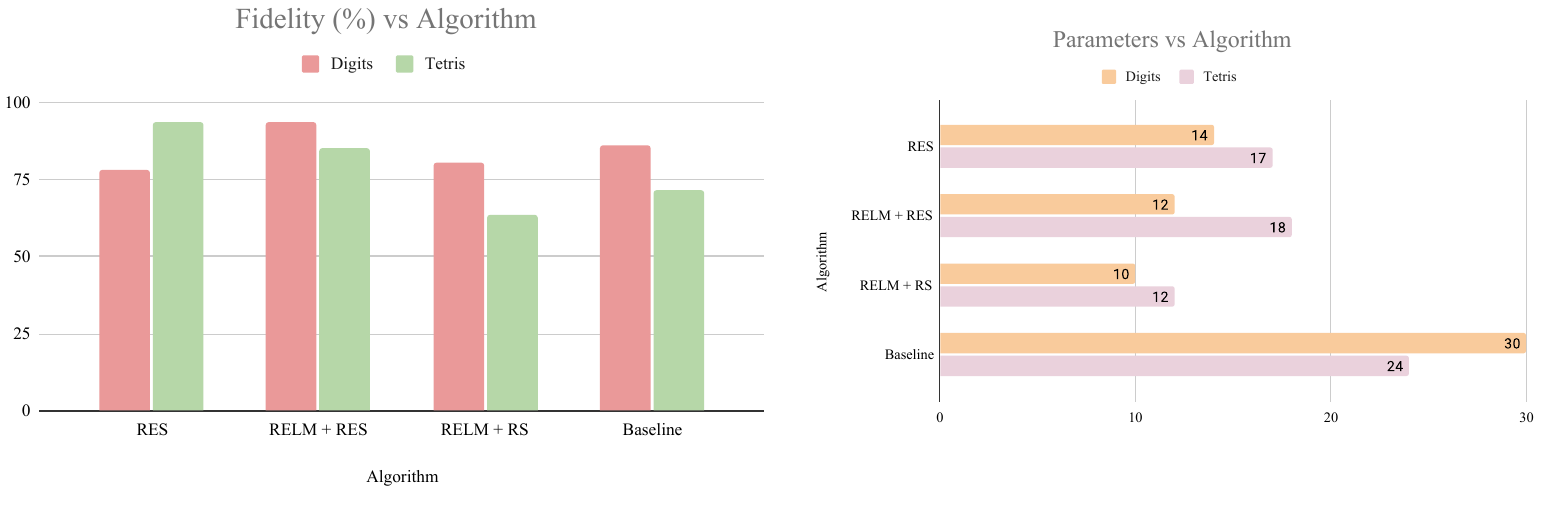}
    \caption{Fidelity and parameter number comparison between baseline and circuits discovered by our proposed search algorithms. The fidelity is estimated as a mean of five independent evaluations on the test set,}
    \label{fig:img_compression_results}
\end{figure*}

\subsection{Image Compression with Quantum Circuits}

In this task, we wish to find the quantum circuit that can reliably compress given input image data into a fewer number of qubits. It is common for current quantum machine learning (QML) tasks to reduce the size of the data by using techniques like PCA before training the parameters. However, these techniques often do not capture a good low-rank representation of the data. A quantum procedure on the other hand has the advantage of operating in pure quantum space and hence is less prone to the aforementioned issue.

We evalaute the performance of a QAE in compressing classical data by creating two synthetic image datasets - Digits and Tetris. The first dataset is the Digits dataset which consists of 100 images of the numbers 0 and 1 in an image of size $8 \times 4$. The second dataset is inspired by the work of Liu~\emph{et al}~\cite{liu2021hybrid} and consists 500 images of Tetris blocks divided into four distinct categories. Each image is of the size $4 \times 4$.  For the RES algorithm, we choose the number of layers as the soft constraint and limit the circuit to be at most $3$ layers. We use the same constraints when using RES with the RELM algorithm. 

\begin{figure}[t]
    \centering
    \includegraphics[width=.75\columnwidth]{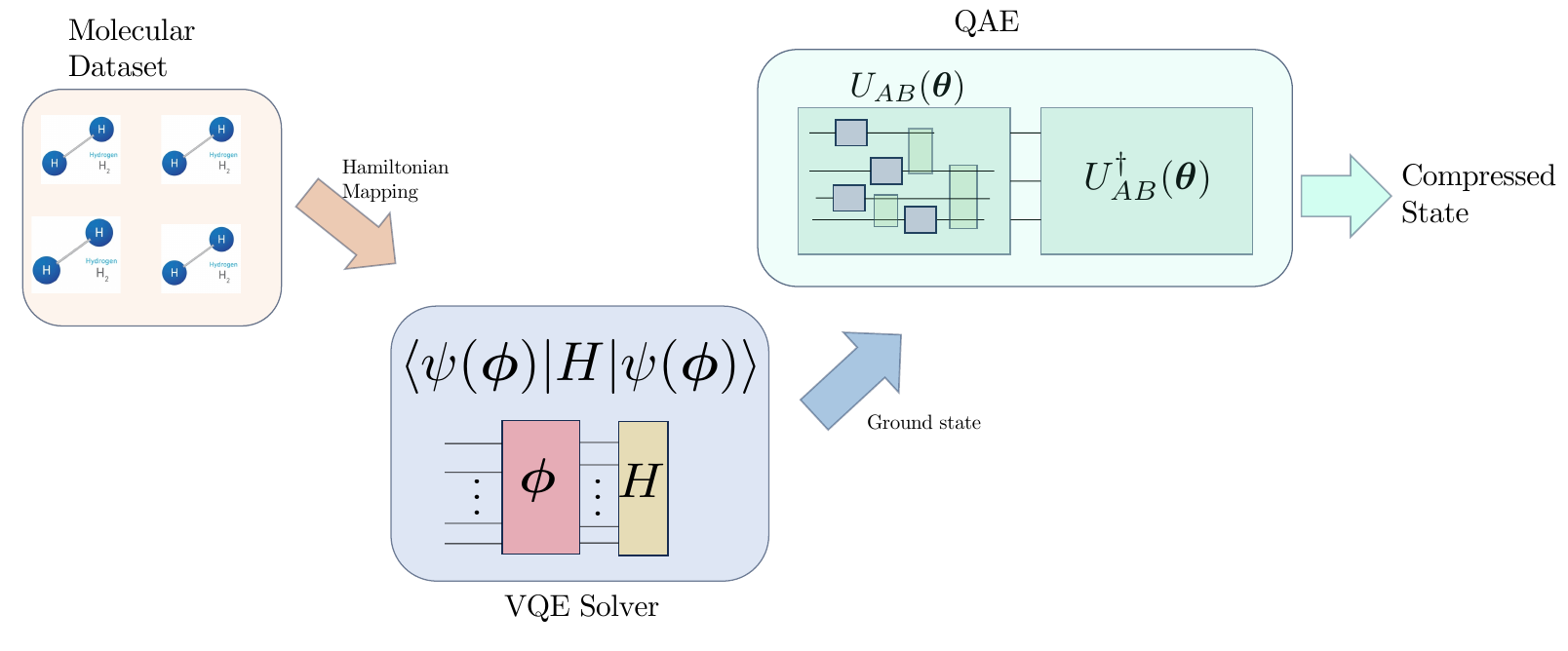}
    \caption{Pipeline for setting up a pure quantum experiment. The dataset is generated by running a variational eigensolver for different bond lengths of a $H_2$ molecule.}
    \label{fig:vqe_setup}
\end{figure}


Figure~\ref{fig:img_compression_results} shows the test set fidelities as well as the number of parameters in the best circuits returned by our search algorithm on the two datasets. We benchmark the performance against a hand tuned baseline circuit consisting of $R_Y$ single qubit rotation and CNOT entanglement gates arranged in 5 layers. The optimal number of layers was found by an incremental search process until a good performance was obtained from the circuit. The fidelities are reported as average of five independent runs of the trained circuits on the test set. We can see that for the Digits dataset, our algorithms yield circuits that have relatively low parameters and a high average fidelity when compared against baseline. More specifically, the RELM algorithm with RES initialization significantly outperforms the baseline. The Tetris dataset has more variance in between the datapoints and consequently it is harder to compress. Even so, we can see that the RES algorithm yields a circuit that has a high compression fidelity when compared against the baseline. As before, we note that the RS initialization fails to provide a significantly diverse population to enable RELM in generating a higher fidelity circuit.

\begin{figure*}[ht]
    \centering
    \includegraphics[width=\columnwidth]{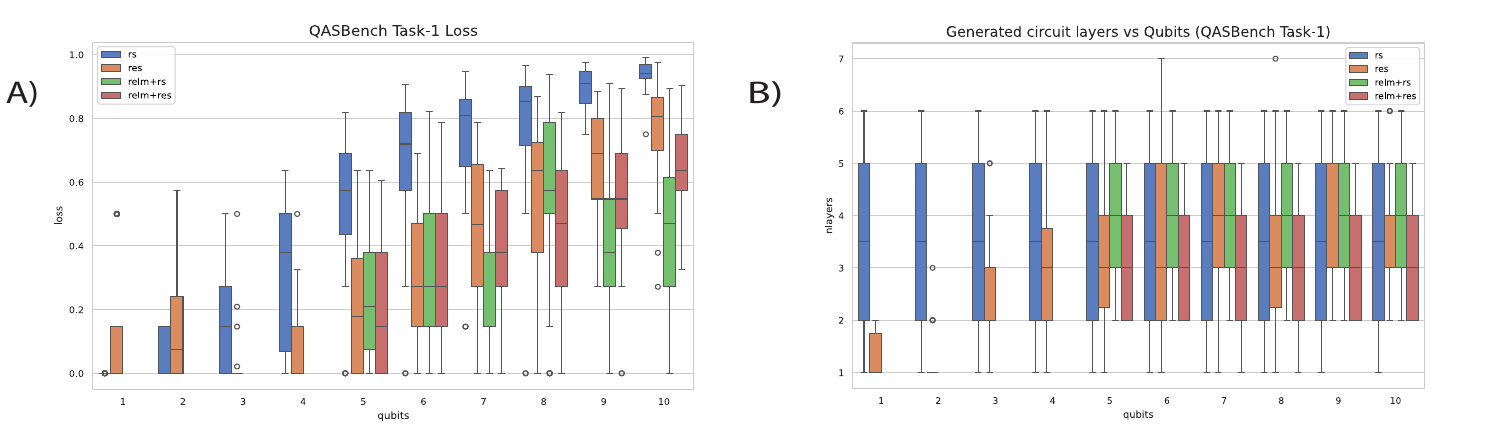}
    \caption{Results of our proposed QAS algorithms on Qasbench task-1. (A) Shows the loss averaged over all circuits for a given number of qubits. (B) Shows the average number of layers in the circuits found with different QAS algorithms and their settings.}
    \label{fig:qasbench_results}
\end{figure*}

\section{Experiments on QAS-Bench Dataset}\label{sec:qasbench}

In order to show the general utility of our algorithms, we perform analysis on standardized data provided by QASBench~\cite{lu2023qas} dataset. We perform experiments on the "Unitary Regeneration" task (i.e. task 1). This task consists of finding the best circuit corresponding to an input unitary matrix. The number of qubits in this task range from $1$ to $10$ and for each qubit there are three types of subtasks (\texttt{dense}, \texttt{hybrid} and \texttt{single}) consisting of unitaries produced by (unknown) circuits with layers ranging from $1$ to $6$. The search space for the first two subtasks is the Clifford set while the last subtask consists of only single qubit Clifford gates.

The authors in~\cite{lu2023qas} provide a ``unitary loss" function that is not efficiently translatable in our algorithms. We thus convert the loss function by evolving $\ket{\psi}_{in} = U_{in}\ket{0}^{\otimes n}$ and measuring $\mathcal{L} = 1 - \langle \psi_g | \psi_{in} \rangle^2$ where $\ket{\psi_g}$ is the state generated by a quantum circuit searched by our QAS algorithm. This loss function is a good proximal function since the unitary loss and fidelity have a symmetric relationship (~\cite{lu2023qas} (Figure 3)). As a baseline, we perform random search (RS) with the given qubit and layer configuration without any additional constraints. This type of search is equivalent to a brute force search across the given search space. We chose this algorithm as our baseline since it is easy to implement, inherently parallelizable and can often find high performing circuits on the given task. For qubits $1 \mapsto 4$ we only run RES to compare against the baseline since we found RELM to be unstable with it's parameter heavy mutation controller for these qubit sizes. We run all three algorithms for qubits ranging from $5 \mapsto 10$. For RES we use the layer constraint as a soft constraint i.e. $g(\mathcal{S}, \alpha) \simeq |L| \leq \tau_l$. For RELM we choose a batch size of $32$ with a learning rate of $3e^{-4}$ and random number generator seed of $6090$. The mutation controller comprises of stacked multihead attention~\cite{vaswani2017attention} layers and is trained with the Adam~\cite{kingma2014adam} optimizer for $30$ epochs. Since the task is different than finding the best encoder in this case, we use a simplified reward function for training the mutation controller with Equation~\ref{eq:vpg_loss}: $R(\tau) = tan(\alpha * (\mathcal{L}_{child} - \mathcal{L}_{parent})*\frac{\pi}{2})$ with $\alpha = 1.5$. 

Figure~\ref{fig:qasbench_results} shows the results of our experiments on the dataset. We profile the loss over all circuits of a given qubit size (Fig~\ref{fig:qasbench_results}(A)) and the number of average number of layers discovered by our algorithm (Fig~\ref{fig:qasbench_results}(B)). It is clear from the loss profiling that our proposed algorithms significantly outperform the baseline. For qubits $1 \mapsto 4$ we see that RES algorithm discovers circuits that either match or outperform the circuits found by random search. The performance of RES monotonically improves over random search even for larger qubits sizes $5 \mapsto 10$. The clearest point of comparison is for the cases of $9$ and $10$ qubit sizes where RES outperforms random search by $21\%$ and $19\%$ respectively.

RELM (with either initialization) significantly outperforms the baseline and RES method. In fact, for $9$ and $10$ qubits, it outperforms baseline by $60\%$, $49\%$ and RES by $49\%$, $38\%$ respectively. Within the different settings however there are some diferences in performance. For instance, RELM with RES initialization works better for qubit sizes $5 \mapsto 8$ while for qubits $9, 10$, RELM with random search initialization seems to perform somewhat better than the RES search. Our hypothesis for this phenomenon is that random search initialization performs more exploration of the search space thereby returning a richer set of parent population. This diverse set can be helpful since for large qubit sizes and layers exploration can prove to be beneficial. On the other hand, RES performs a constrained search which limits it to finding architectures that perform better within a particular budget. This type of search then can yield somewhat less diverse initial set of parent cells to the algorithm.

From Figure~\ref{fig:qasbench_results} (B) we can also see that random search often finds circuits with large number of layers. Since we use a layer constraint in the RES search for this task, we find that RES often generates circuits that outperform random search and still are more compact than the ones found by random search. This trend also persists when RELM is used with RES initialization. These results lead us to an important realization - we can indirectly use soft constraints in RELM as well by simply initializing with RES and our chosen soft constraint. This combination can be extremely useful in cases where finding circuits within some budget is highly critical. We present additional results in Appendix~\ref{sec:qasbench_more_res}.

\begin{table*}[t]
    \centering
    \small
    \begin{tabular}{cccccc}
    \toprule
    Method & Global Search & Local Search & Soft Constraints & Learned Mutation & Parallelizability \\
    \midrule
    Reinforcement Learning & \ding{51} & \ding{51} & \ding{55} & \ding{51} & \ding{55} \\ 

    Evolutionary Search & \ding{51} & \ding{51} & \ding{55} & \ding{55} & \ding{51} \\ 

    Differentiable QAS & \ding{51} & \ding{51} & \ding{55} & \ding{55} & \ding{55}\\

    \hline 

    \textbf{RELM/ RES} & \ding{51} & \ding{51} & \ding{51} & \ding{51} & \ding{51}
    \end{tabular}
    \caption{Algorithmic differences between major categories of QAS algorithm methods and our proposed algorithms.}
    \label{tab:relwrk}
\end{table*}

\section{Generalization to Other Quantum Tasks}\label{sec:generalizability}
In this section, we discuss the broader applicability of our method to other quantum algorithms. Specifically, we analyze how our proposed methods can be used with such algorithms as the Quantum Approximate Optimization Algorithm (QAOA), Quantum Neural Networks (QNNs) and Variational Quantum Eigensolver (VQE). Note that each of these cases represents a distinct project on its own, and we therefore identify them as promising quantum algorithm-oriented research directions for future exploration.

\subsection{QAOA}
QAOA~\cite{farhi2014qaoa_for_bounded} algorithms typically aim to solve  combinatorial optimization (CO) problems whose objective function can be encoded into a Hamiltonian. The general framework of a QAOA circuit is an alternating application of the ``cost'' and ``mixer'' operators. The former is driven by the type of combinatorial optimization problem being solved. For example, if one is solving the graph maximum cut problem then the cost operator minimizes the cost 
\[
C(z) = -\frac{1}{2}\sum_{(u, v) \in \mathcal{E}} (1 - \sigma_{z_u} \sigma_{z_v}),
\]
where $\mathcal{E}$ is the set of graph edges. Given cost and mixer parameters $\bm{\gamma}, \bm{\beta} \in \mathbb{R}^p$ ($p$ is the number of repetitions of the alternating cost and mixer operators, also known as the depth of QAOA), a QAOA circuit minimizes $J(\bm{\gamma}, \bm{\beta}) = \bra{\bm{\gamma}, \bm{\beta}} C(z) \ket{\bm{\gamma}, \bm{\beta}}$. 

We can leverage our proposed QAS algorithms to search for mixer circuits that provide a lower cost than hand-designed circuits on the given combinatorial optimization problem. Another major bottleneck in QAOA applications is the number of repetitions $p$ needed to achieve a low cost for a given objective function. This is also directly related to the hardness of classical optimization component of the QAOA variational cycle in which $\bm{\gamma}, \bm{\beta}$ are optimized. Many techniques have been suggested to overcome this bottleneck and accelerate this optimization \cite{falla2024graph,galda2023similarity}. We can directly apply RES to a given QAOA problem with the soft constraint being a budget on $p$, i.e. $g(\mathcal{S}, \alpha) \simeq p \leq \tau_p$. Once some candidate circuits have been found, we can leverage RELM to further reduce the cost by finding mutations that lead to improved performance with an even lower depth. A key advantage of our algorithms here is that we do \emph{not} assume anything about the cost function. Hence, our method is task agnostic and generally applicable to QAOA problems.

\subsection{QNN}

Quantum Machine Learning (QML)~\cite{schuld2015introduction} typically involves solving problems such as classification or regression by using quantum versions of classical machine learning models. The most popular models are QNNs that aim to train parameters $\bm{\theta}_{qnn}$ such that a cost function $C(\bm{\theta}) = \bra{\psi(\bm{\theta})}O\ket{\psi(\bm{\theta})}$ is minimized.  Here $\ket{\psi(\bm{\theta})} = U(\bm{\theta})\ket{\psi_x}$ and $O$ is a given observable. There are two main components in a typical QNN: a data embedding and a task circuit. The former circuit for data embedding loads classical data into a quantum state. For example, an amplitude embedding circuit builds an $n$-qubit quantum state as
\begin{equation*}
    \ket{\psi_x} = \sum_{i=1}^n x_i \ket{i},
\end{equation*}
where $x_i$ is the i$^{th}$ component of the input data point $\bm{x}$. Finding optimal embedding circuits is not a well defined QAS problem. However, finding an efficient task circuit is highly suitable for QAS algorithms. In QML setting, the biggest bottleneck is the total number of parameters since training a QNN requires computing a gradient 

\begin{equation*}
    \frac{\partial C(\bm{\theta})}{\partial \theta_k} = \frac{1}{c}[C(\theta_k + s) - C(\theta_k - s)],
\end{equation*}
where $C(\theta_k \pm s)$ indicates addition/subtraction with the $k^{th}$ parameter component. This computation has a linear scaling in the total number of parameters and consequently training a QNN has \emph{quadratic} cost. Hence, an efficient task circuit is one that achieves low cost with less number of parameters. We can easily leverage RES in this case to search for circuits within a specified parameter budget i.e. $g(\mathcal{S}, \alpha) \simeq |\bm{\theta}| \leq \tau_{params}$. The RELM algorithm can then be used to further reduce the cost by finding mutations that consume even less parameters than the total budget while maintaining or exceeding the performance of circuit discovered by the RES algorithm. 

\subsection{VQE}
The VQE~\cite{VQE} algorithm was proposed to solve quantum chemistry problems with variational circuits. It aims to find the minimum energy of a given chemical molecule by minimizing cost function of the form:

\begin{equation*}
    C(\bm{\theta}) = \bra{\psi(\bm{\theta})}H_{mol}\ket{\psi(\bm{\theta})}.
\end{equation*}

The minimum of this cost function corresponds to the upper bound on the ground state energy $E$ of a given molecule. The trial wave function, $\ket{\psi(\bm{\theta})}$, is prepared through a variational circuit applied to an input ``Hartree-Fock'' state. In this task, the quality of ansatz directly dictates the performance of the eigensolver. Furthermore, the VQE algorithm is also a gradient based algorithm. Hence, an efficient ansatz in this task must not only achieve lower cost than a hand-designed circuit~\cite{VQE} but also must consume lesser parameters. 

Our proposed algorithms are ideally suited for this task. Similar to QNN, we can set the soft constraint on the total number of parameters on the circuit and run RES algorithm to obtain good initial candidates with parameter budget. Then RELM can be run to refine the structure of these circuits such that it outperforms a human-designed circuit.

\subsection{Discussions on Scalability}

One of the key advantages of our proposed algorithms is that they are engineered towards being scalable from ground up. We achieve this by two primary mechanisms. The first mechanism is information reuse. RES incrementally searches a good candidate circuit by utilizing information from simpler and shallower circuits. Similarly, RELM also exploits information provided by RES to simplify a candidate circuit in order to satisfy the constraints of the search. The second mechanism is parallelizability. Specifically, RES algorithm utilizes parallel search with seed circuits. Additionally, RELM can also run mutation pipeline in parallel across multiple GPUs to efficiently search for the mutation policy.

As a concrete example, let's consider the action of a conventional algorithm like QuantumDARTS~\cite{wu2023quantumdarts} with our proposed algorithms. In QuantumDARTS, the macro-search algorithm samples different circuits according to the search criteria \emph{independently} at each ``epoch" of it's run. Thus, any information gained from previous epochs is discarded. In contrast, RES efficiently re-uses information from previous searches to intelligently look for better circuits. Similarly, micro-search algorithm in QuantumDARTS also discards information during it's run. Additionally, QuantumDARTS search algorithms are not easily paralleizable. Each step of the algorithm not only finds a structure of the circuit but also the optimal parameters. These two steps are sequential in nature and thus cannot be run in parallel. We further argue that finding optimal parameters is not a necessity of a quantum search algorithm. Finding \emph{structure} that fits a user's constraints on a given class of problems is far more useful and relevant. In this respect, the parallel nature of our algorithms can provide a user with a good performing structure that fits within their budget constraints faster than a sequential search algorithm like above. 

To further look at scalability in terms of qubit sizes, we would like to draw the reader's attention to Figure~\ref{fig:qasbench_results}. These experiments show that our algorithms are able to handle problem sizes larger than the ones considered in our main experiments. Specifically, we show that our algorithms can handle circuits of large qubit sizes and still find circuits within a reasonable depth budget. The qubit sizes considered in the QasBench experiments either match or exceed the qubit sizes considered in related works~\cite{wu2023quantumdarts, bilkis2021semi, duong2022quantum}.

Combining insights from qualitative and quantitative comparisons above allows us to conclude that our algorithms are more scalable than existing works in literature owing to information reuse and parallelizability.

\section{Related Work}\label{sec:rel_wrk}
We now compare our proposed algorithms with other QAS algorithms proposed in literature. We categorize existing QAS algorithms by their algorithmic themes and compare their relative advantages and disadvantages with our method. Table~\ref{tab:relwrk} summarizes our analysis by comparing various methods used in QAS algorithms on different characteristics.

\subsection{Reinforcement Learning Methods}
Reinforcement learning (RL) methods aim to generate an optimal circuit by casting the QAS problem as a Markov Decision Process (MDP). An MDP is defined by the tuple $(S, \mathcal{A}, \mathcal{R}, \gamma)$ where $S$ is the state space. In the context of QAS algorithms, the state space comprises of various combinations of supported gates in the given search space $\mathcal{S}$. The action space $\mathcal{A}$ is usually a discrete set that dictates the placement of a supported gate on a qubit. $\mathcal{R}$ and $\gamma$ are reward and discount factors respectively. The overall goal is to train a policy $\pi(\bm{\phi}): S \mapsto \mathcal{A}$, such that optimal parameters $\bm{\phi}*$ correspond to the optimal policy $\pi*$. The parameters are trained using well known optimization algorithms such as  REINFORCE~\cite{sutton1999policy}, and  PPO~\cite{schulman2017proximal}.

Many different RL methods have been proposed for QAS algorithms in the literature for different types of quantum tasks. Ostaszewski~\emph{et al}~\cite{ostaszewski2021reinforcement} propose to use a combination of RL and curriculum learning to search for efficient quantum circuits for VQE task. Kundu~\emph{et al}~\cite{kundu2024enhancing} propose to use Double Deep Q-Network for variational quantum state diagonalization task. McKiernan~\emph{et al}~\cite{mckiernan2019automated} have also proposed to use RL algorithms for finding circuits for combinatorial optimization problems like MaxCut, QUBO etc. 
Although RL has shown impressive success in machine learning, its application in the QAS setting comes with challenges. One of the most significant is the large amount of data typically required for the algorithm to learn an optimal policy 
(e.g. in~\cite{mckiernan2019automated}, it took 1,700,000 episodes to generate an optimal policy). Secondly, an RL algorithm is very hard to parallelize efficiently workers are required to send vector valued updates. As the size of the neural network grows, the communication costs between various workers outweigh any potential advantages gained by parallelization.

\subsection{Evolutionary Algorithms}
Evolutionary algorithms (EA)  approach search in a different way. There are three main components in an EA, namely, the data to phenotype encoding, genetic operators and a fitness function. The last component is similar to the reward function in RL.  Various different EA optimizers (e.g. CMA-ES~\cite{hansen2016cma}, CrossEntropy  method~\cite{rubinstein2004cross} or tournament selection) can be used to optimize an initialized ``population'' of given quantum circuits' encodings to gradually discover the one that has the best value of the given fitness function. 

The QAS setup has a direct mapping on the main components of the evolutionary algorithm. This is why it is another popular choice for many QAS methods in literature. Mog-VQE~\cite{chivilikhin2020mog} and EVQE~\cite{rattew2020evoansatz} considered evolutionary algorithms for VQE task. It must be noted that while Mog-VQE uses a multi-objective approach similar to RES, their algorithm optimizes for the different objectives sequentially and thus tries to find a \emph{global} ansatz satisfying the constraints. In contrast, our method can \emph{simultaneously} optimize for given constraints while generating non-intuitive circuit structures. Other methods include QCEAT~\cite{huang2022robust} which uses a multi-species approach to find resource efficient circuits. 

The EA methods are broadly applicable and easy to implement. Despite their relative advantages, they do have a few drawbacks. One major drawback is that EA methods always try to seek the global optimum given a fitness function. In cases when the fitness function is not well defined, the search process can take a long time to converge. Another drawback is that the performance of EA is conditioned on the encoding strategy. Since this varies highly from application to application, the performance of EA needs to be evaluated for different styles of encodings for a given task. This often leads to a longer development time than other algorithms.

\subsection{Differentiable QAS Methods}
In the methods presented thus far, the search space $\mathcal{S}$ is discrete. A search algorithm then is restricted to sampling discrete combinations from this search space and evaluating possible architectures. Owing to the discrete nature of $\mathcal{S}$ it is impossible to create a differentiable loss function of the form $\mathcal{L}: \mathcal{S} \mapsto \mathbb{R}$. Having access to this type of loss function is beneficial since we can evaluate the ``goodness" of a candidate architecture sampled from $\mathcal{S}$ by simply evaluating the gradient of the loss function. Differentiable QAS methods provide such a framework by making the search space continuous with a proxy variable $\bm{\alpha} = \{\alpha_{s_1}, \alpha_{s_2} \dots \}$ that assigns a score $p(\alpha_s)$ to a given architecture $s \in \mathcal{S}$. The best architecture is the one that corresponds to the best score $\alpha^* = \argmax_\alpha \mathcal{L}(\alpha)$.

DQAS~\cite{zhang2020differentiable} considers a differentiable strategy in the spirit of DARTS~\cite{liu2018darts} from classical NAS algorithms for QAOA tasks. The calculate the gradient $\nabla_\alpha \mathcal{L}(\alpha)$  via Monte-Carlo sampling. Following this work, QuantumDARTS~\cite{wu2023quantumdarts} was proposed recently. They propose two variants of differentiable search  - macro and micro. The former is concerned with generating a quantum circuit up to arbitrary depth given a cost function. The latter is concerned with generating a good performing layer which is then stacked up to arbitrary depth. The gradient $\nabla_\alpha \mathcal{L}(\alpha)$ is calculated via the Gumbel-Softmax distribution. We note that the micro search strategy of QuantumDARTS is similar to our proposed methods. A key difference between these methods is that we generate architectures that offer best performance by searching for best performing layer at each expansion/mutation while their method only generates a single optimized architecture and then builds an ansatz by stacking it to a depth.

Despite their attractive properties, differentiable QAS methods are not without their drawbacks. The first drawback is that they are computationally expensive - Monte Carlo sampling of gradients is not sample efficient. The second and more serious drawback is the dependence of these methods on the latent variable $\alpha$ to characterize the search state. Owing to this latent dependence, modeling multiple objectives with these methods requires significant changes to ensure candidate architectures are scored according to those objectives. The third drawback is their lack of parallelizability. This defect arises due to the need for keeping track of the scores of \emph{all} candidate solutions generated during the forward pass to allow for proper gradient computation. 


\section{Conclusions}\label{sec:conclusion}

We have presented a new way of looking at the problem of Quantum Architecture Search. Previous works in this domain have mainly tried to apply the techniques that exist in the NAS literature and have thus focused on trying to find a global optimal circuit that can be used for a given dataset and task. On the other hand, we presented two approaches that focus on finding \emph{quantum circuit patterns} that can increase performance even when the number of parameters is not high enough. Moreover, our proposed algorithms are the first to allow for incorporating \emph{resource constraints} directly in the search procedure resulting in efficient circuits. The experimental results demonstrate that we improve upon existing architecture search approaches for quantum circuits by generating architectures that provide a high performance on the provided metric while consuming less resources than other approaches. Moreover, our approaches stand out for being \emph{customizable} in terms of the specific type of resources a user may want to preserve. 


Several research directions emerge from our work. One possible direction is to apply the proposed NAS algorithms for combinatorial optimization problem and compare against existing methods~\cite{kulshrestha2023qarchsearch}. In the RELM algorithm, we have used the vanilla policy gradient loss for training the mutation network. Interesting extensiond to this algorithm would be to train the mutation network using an actor-critic framework~\cite{mnih2016asynchronous} and using it within quantum mixture of experts \cite{nguyen2025qmoe}. In this work we have exclusively dealt with searching for quantum architectures. However, a novel application of our work can focus on generating efficient architectures for large language models that can provide the same fidelity as original models but consume a fraction of resources required to run them at scale.


\section*{Acknowledgements}

This work was partially supported with funding from the Defense Advanced Research Projects Agency (DARPA) under the ONISQ program and by National Science Foundation award \#2444042.

\appendix
\section{RELM Mutation Pipeline}\label{sec:relm_mut}

\begin{figure*}[t]
    \centering
    \includegraphics[width=\columnwidth]{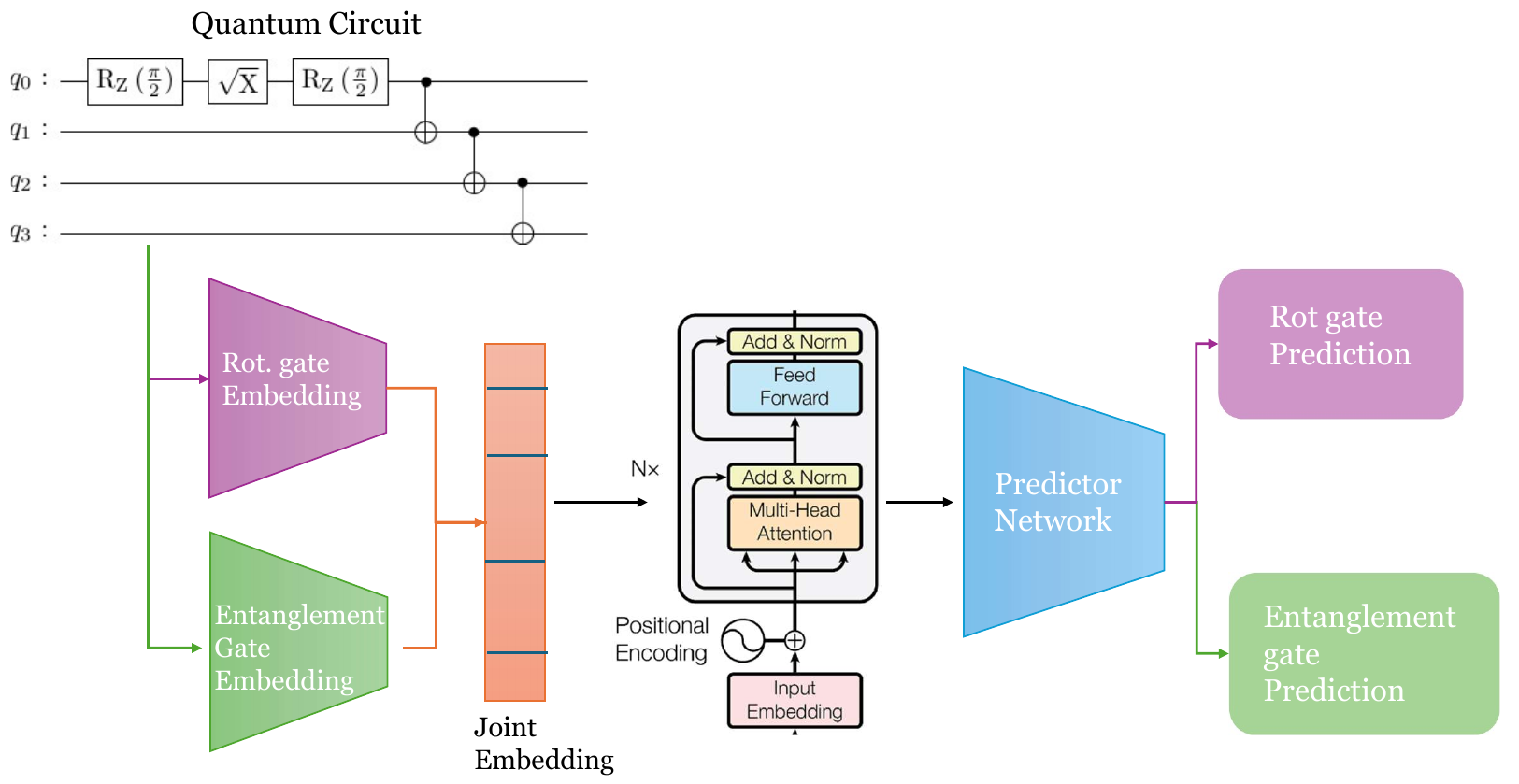}
    \caption{RELM Circuit Mutation Pipeline. We use transformer 
    encoders to process circuit ``views" and predict the next mutation}
    \label{fig:relm_enc_flow}
\end{figure*}

Figure~\ref{fig:relm_enc_flow} shows the hybrid quantum-classical pipeline for learning a reinforcement learning policy $\pi_{\phi}$ that learns to generate mutations that outperform the input circuit in terms of a given performance metric (e.g. fidelity of recovered quantum state after compression). In broad strokes, there are three main steps in the pipeline: conversion from a quantum representation to classical, 
a policy update step and conversion of classical representation back to quantum representation. 

The first step in the pipeline is to convert a quantum representation into a classical representation which can be used to train a deep neural network. In order to acheive this conversion, we first create an intermediate representation (IR) of a quantum circuit that we call a \texttt{cell}. The \texttt{cell} is a dictionary that holds a key-value pair of every single and multi-qubit operation in the circuit. We broadly categorize these key value pairs into rotation and entanglement operations. The former maps operations that operate on a single qubit (e.g. parameterized rotation along an axis). The latter maps operations that involve more than one qubit (e.g. CNOT gates). All operations are mapped to an integer in a ``gate vocab". We use this gate vocab to build one-hot vectors of different operations. The individual vectors belonging to a particular category of operation are then concatenated into a tensor. 

The second step is to train the neural network. We have already outlined the sizes of different matrices that are used to generate embeddings for the rotation and entanglement operations in earlier sections. Once the joint embeddings are obtained they are processed through $N$ transformer encoders. These encodings are then projected back to the respective rotation and embedding embedddings. We apply a softmax at the end of the output projections for each category of operations and select the best operation to apply at every qubit. The network is trained using vanilla policy gradient with the reward function described in the main text. 

The final step is to convert the recommended operations back to a possible quantum circuit. In order to achieve this, we take the index of the recommended operations and create the IR as before. Our \texttt{cell} class then takes the responsibility of converting the new operations into a valid quantum circuit by performing an inverse operation from the previously built gate vocab. We note that this transform from classical to quantum does not produce a unique circuit since we don't encode or enforce the \emph{order} of operations in our IR. Since our investigation is to generate many possible quantum circuits for a given application, we believe this is a reasonable trade-off between computation complexity and effectiveness of our algorithm. However, in cases where order is important (e.g. quantum compilation tasks) our IR can be easily extended to incorporate this feature as well.

\section{Experimental Setup Details}

In this section, we describe in details the image datasets and the baseline circuits used in the hybrid quantum setting of our experiments which is required for reproducibility and experiments.

\begin{figure*}
    \centering
    \includegraphics[width=\columnwidth]{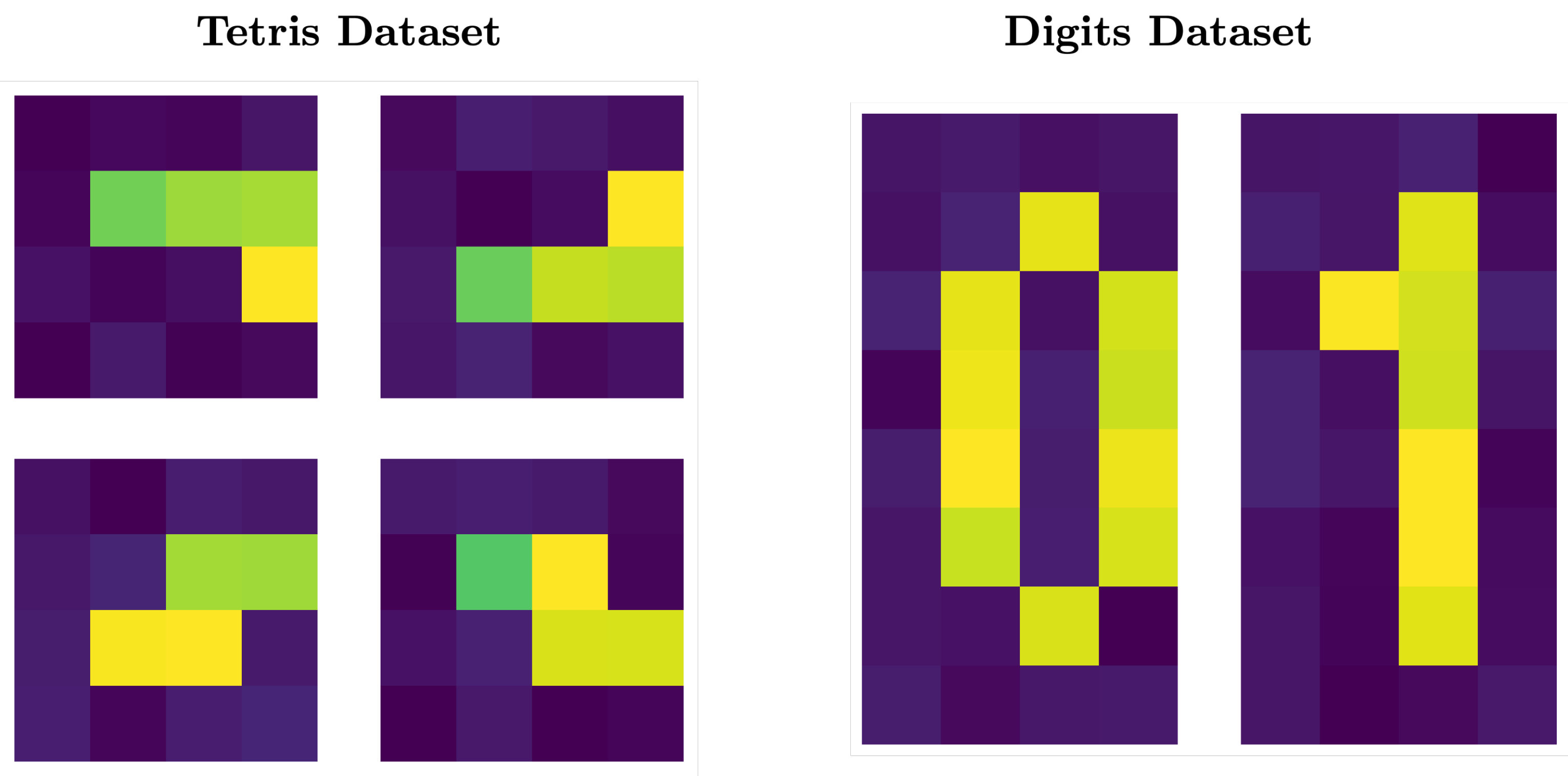}
    \caption{Representative images of the datasets used in the hybrid quantum setting of our experiments.}
    \label{fig:img_datasets}
\end{figure*}

Figure~\ref{fig:img_datasets} shows the representative images used in the image compression experiments. The Tetris dataset was inspired by the work of Liu~\emph{et al.}~\cite{liu2021hybrid}. However, we use an image size of $4 \times 4$ and restrict ourselves to four types of Tetris blocks. The digits dataset is a simplified version of the MNIST dataset where the background pixels are chosen uniformly between $[0.01, 0.05]$ and the foreground pixels are uniformly. chosen from $[0.5, 1]$.

\begin{figure*}[h]
    \centering
    \includegraphics[width=\columnwidth]{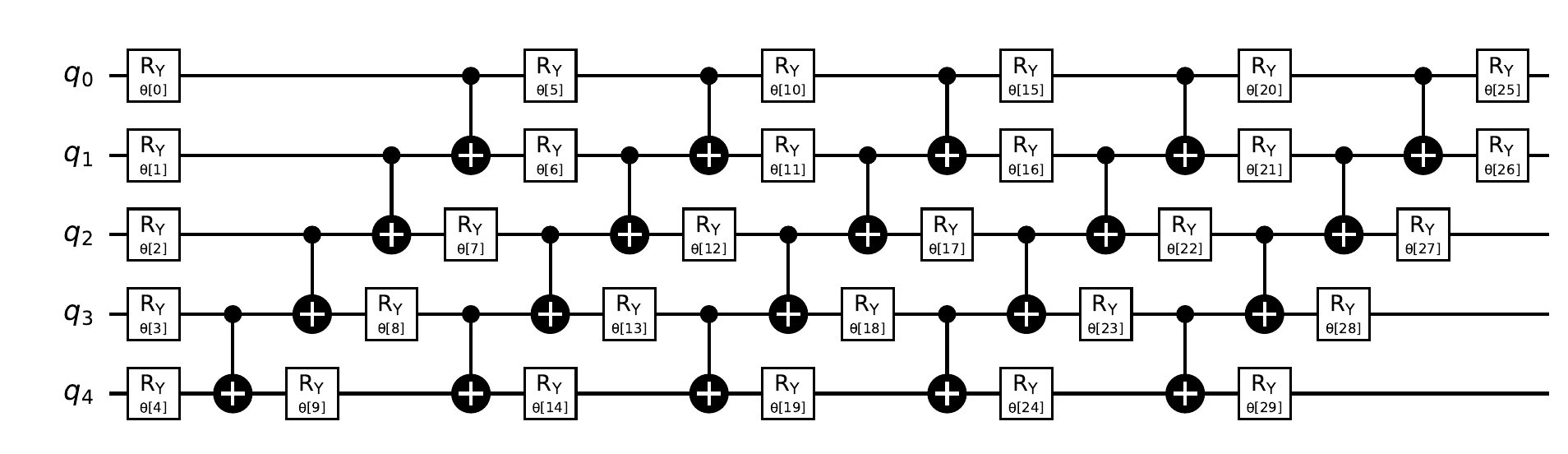}
    \caption{Baseline encoder for the QAE model in the image compression datasets}
    \label{fig:baseline_ae}
\end{figure*}


In order to benchmark our image compression performance on these datasets we created a simple baseline circuit shown in Figure~\ref{fig:baseline_ae}. It comprises of alternating layers of $R_Y$ rotation and $CNOT$ entanglement gates. For the case of Digits dataset we encode the image into $5$ qubits and for the case of Tetris dataset we encode the image into $4$ qubits using Amplitude encoding:

\begin{equation}\label{eq:amp_embed}
    \ket{\psi_x} = \sum_{i=1}^{N} x_i \ket{i}
\end{equation}

Here $x_i$ is the $i^{th}$ pixel value in the image vector and $\ket{i}$ is a basis vector (a one hot vector that is 1 at index $i$). The number of qubits $N = log_2(n)$ where $n$ is the flattened size of the input. In our code, we use \texttt{RawFeatureMap} provided by the Qiskit library to accomplish this encoding.


\section{Discovered circuits}\label{sec:disc_ckts}

In this section we present examples of the most optimal circuits that were discovered by our algorithm in three different settings - vanilla RES, RELM with random search initialization and RELM with RES initialization. More specifically, we present discovered circuits for image compression and denoising tasks.

\begin{figure*}[h]
    \centering
    \includegraphics[width=\columnwidth]{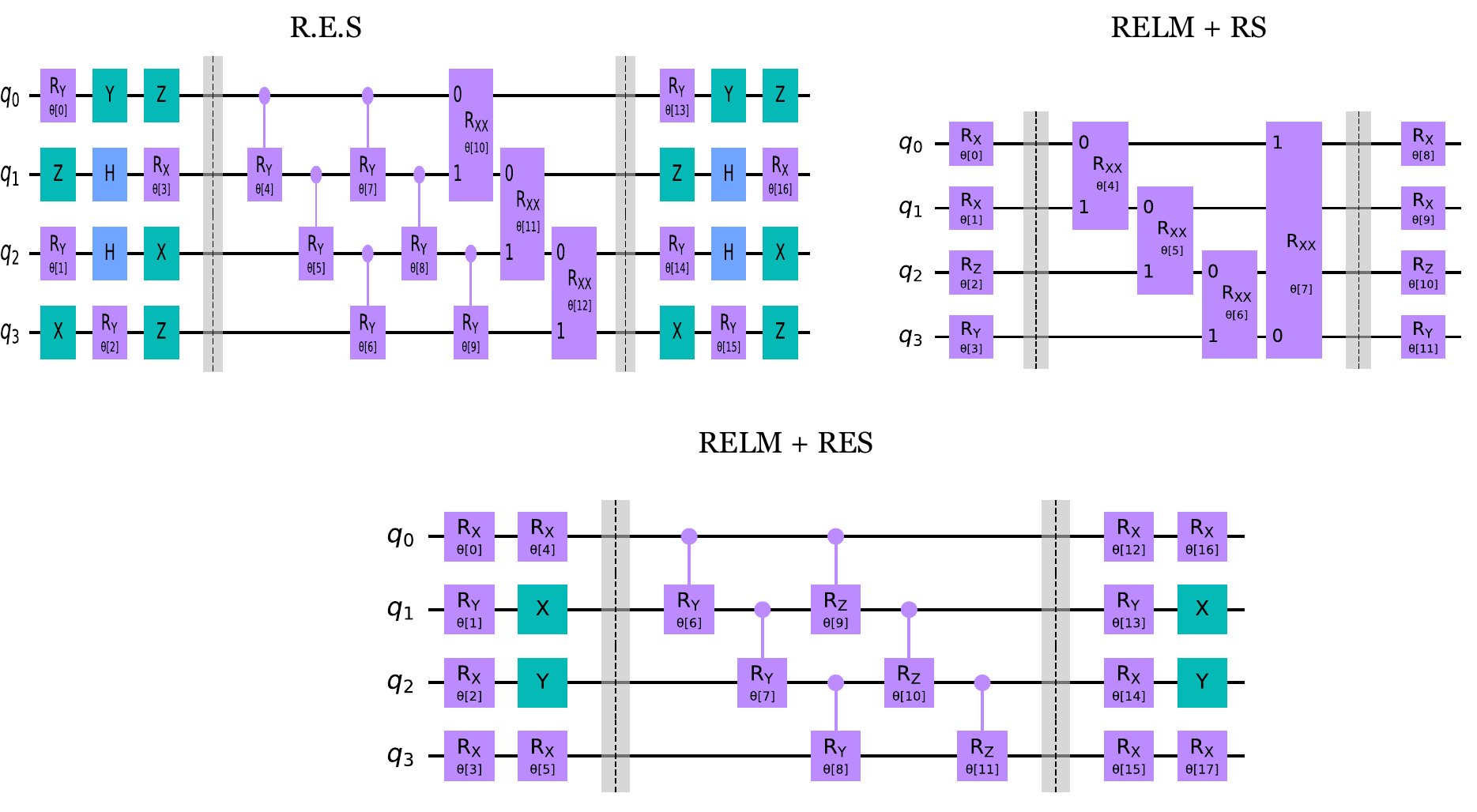}
    \caption{Quantum circuit patterns discovered by the proposed NAS algorithms}
    \label{fig:searched_ckts}
\end{figure*}

\begin{figure*}
    \centering
    \includegraphics[width=\columnwidth]{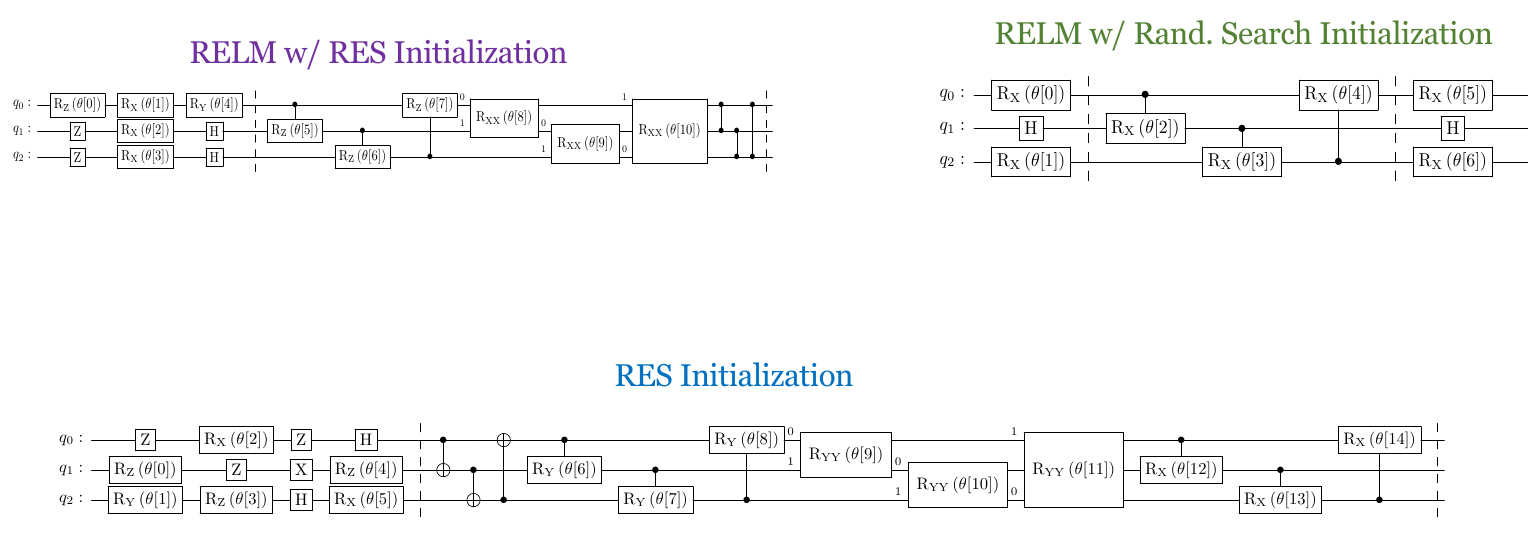}
    \caption{Quantum circuit patterns discovered by our algorithms for denoising bitflip noise corrupted states.}
    \label{fig:searched_ckts_dnbf}
\end{figure*}

Figure~\ref{fig:searched_ckts} shows the best circuits discovered by our proposed algorithms in different combinations for the image data compression task. We can see that all discovered circuits differ significantly from the baseline circuit AE circuit. In the case of RES algorithm (top left) we see that the discovered circuit favors more single qubit rotation gates with a full entanglement. When the RELM procedure is run with randomly generated initial population (top middle), we can see that the algorithm only succeeds in discovering a simple repeatable pattern that fits within the given constraint. This pattern closely resembles the baseline circuit we used in our image compression experiments. However, when RELM procedure is run with the initial population generated by the RES procedure (bottom) we can see that the generated circuit has \emph{controlled complexity} i.e. neither does not have too many gates compared to the circuit discovered by standalone RES algorithm  nor too few gates in the RELM + random search case. As demonstrated in the results, this circuit outperforms all other proposed/hand-designed circuits in our study.

Figure~\ref{fig:searched_ckts_dnbf} shows the best circuits discovered by the different algorithms for the denoising task when bitflip noise is used as the noise source. Even though the circuits are markedly different from the previous applications similar patterns emerge as in the previous circuits. For the standalone RES algorithm (bottom), we can see that the discovered circuits again use a lot of single qubit gates and prefer a fully entangled two qubit gate strategy. The circuit discovered by RELM 
with random search initialization (top middle) also shows a similar preference towards simpler repeatable circuit patterns. Finally, RELM with RES initialization (top left) shows a design with controlled complexity and uses fewer gates than standalone RES. From these examples we can reasonably conclude that RELM when paired with RES can discover non-intuitive circuits that fit any given constraints.

\section{Further QASBench Results}\label{sec:qasbench_more_res}

\begin{figure}[h]
    \centering
    \includegraphics[width=.75\columnwidth]{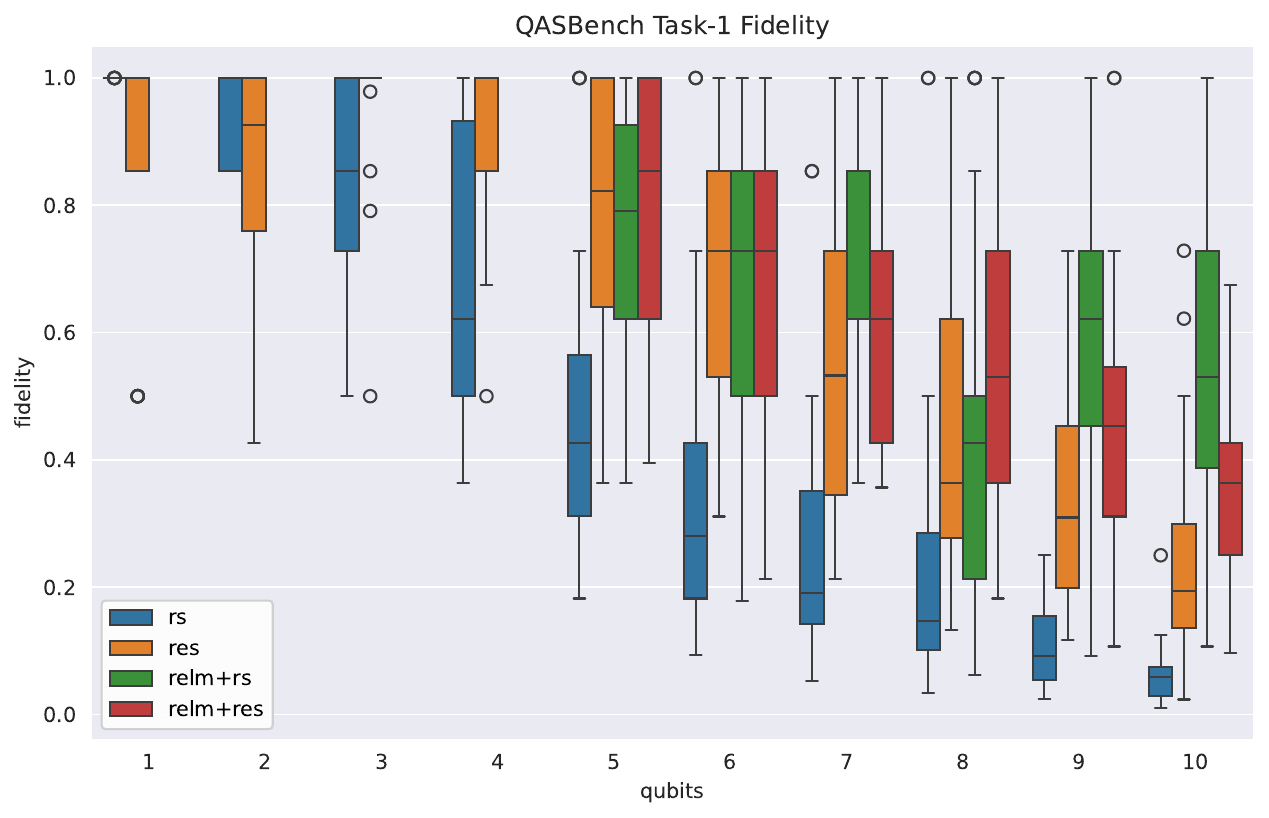}
    \caption{Fidelities of cells discovered with the input state evolved by the given unitary matrix.}
    \label{fig:qasbench_fid}
\end{figure}

To estimate the quality of circuits discovered by our algorithms we measure the fidelity of the output state produced by our circuit with the input state produced by the given unitary matrix. Figure~\ref{fig:qasbench_fid} shows the results of these profiling experiments. We can see the symmetric relation between the loss function and fidelity we alluded to in the main section. Furthermore, the fidelities show less degradation in performance than baseline when the number of qubits is increased.

\begin{figure}[t]
    \centering
    \includegraphics[width=.75\columnwidth]{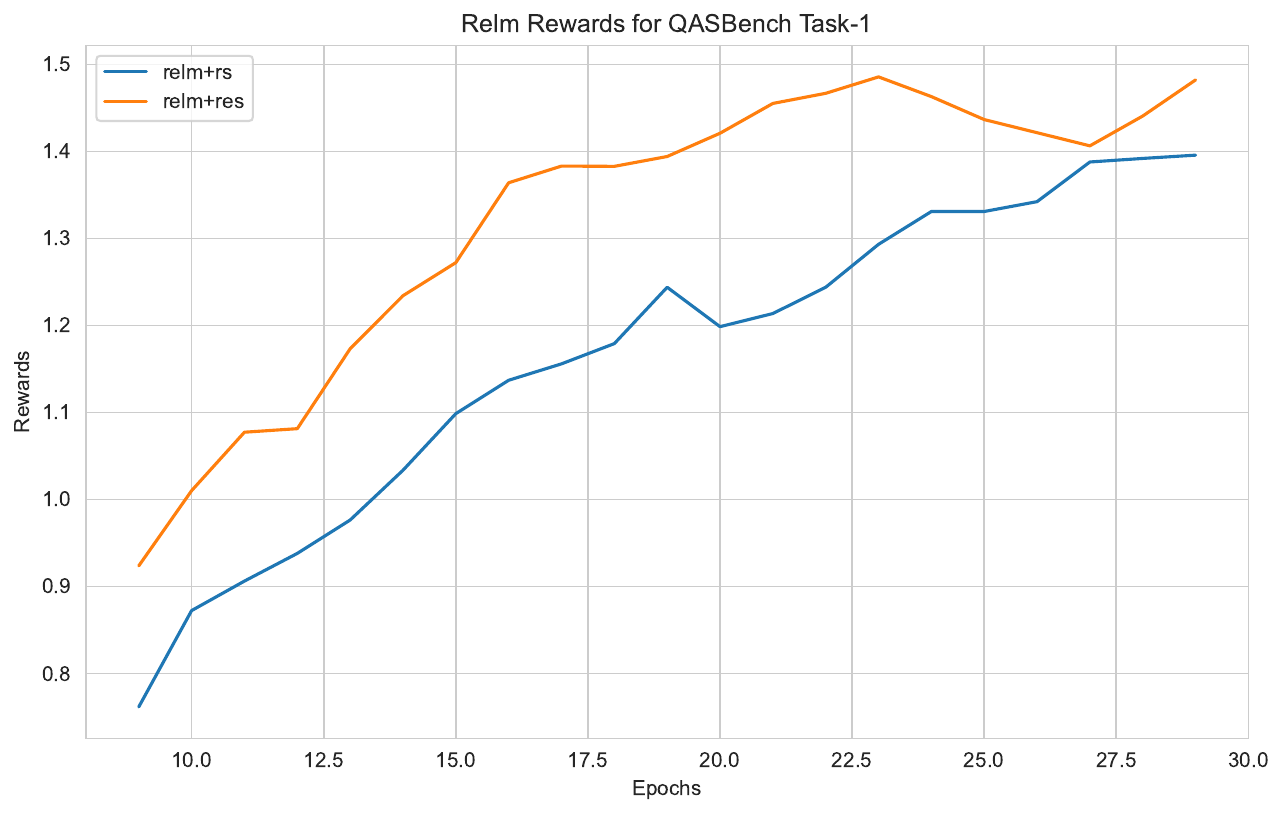}
    \caption{Average RL rewards over 30 epochs of RELM for a given input unitary matrix with random search (RS) and RES initializations}
    \label{fig:rl_viz}
\end{figure}

To further investigate the learning process, we profile the reward obtained by RELM with different initialization strategies on a single input matrix. Figure~\ref{fig:rl_viz} shows the smoothed rewards obtained over the epochs. We keep the seed and learning rate to be exactly the same as the one in our main experiments. In the case of RELM with RS initialization, we see that initially rewards grow slowly (indicating exploration of search space) before accelerating in the middle epochs. On the other hand, since with RES initialization, the initial reward is high indicating strong feedback from conditionally searched cells. We also see that the reward from RES initialization is higher than with RS initialization.

\clearpage

\bibliographystyle{IEEEtran}
\bibliography{quantum_ref}

\end{document}